\PassOptionsToPackage{table}{xcolor}
\documentclass{article}
\usepackage[T1]{fontenc}
\usepackage{antibody_preprint,times}

\usepackage{amsmath,amsfonts,bm}

\def\eqref#1{equation~\ref{#1}}
\def\1{\bm{1}}

\DeclareMathAlphabet{\mathsfit}{\encodingdefault}{\sfdefault}{m}{sl}
\SetMathAlphabet{\mathsfit}{bold}{\encodingdefault}{\sfdefault}{bx}{n}

\usepackage{amsmath}
\usepackage{amsthm}
\usepackage{booktabs}
\usepackage{capt-of}
\usepackage{flafter}
\usepackage{float}
\usepackage{graphicx}
\usepackage{longtable}
\usepackage{makecell}
\usepackage{microtype}
\usepackage{multirow}
\usepackage{needspace}
\usepackage{tabularx}
\usepackage{xltabular}
\definecolor{tableink}{RGB}{42,52,55}
\definecolor{tableaccent}{RGB}{74,106,108}
\definecolor{tableline}{RGB}{194,205,206}
\newcommand{\tabletoprule}{\arrayrulecolor{tableline}\specialrule{0.65pt}{0pt}{0.45em}}
\newcommand{\tablemidrule}{\arrayrulecolor{tableline}\specialrule{0.35pt}{0.35em}{0.35em}}
\newcommand{\tablebottomrule}{\arrayrulecolor{tableline}\specialrule{0.65pt}{0.45em}{0pt}}
\newcommand{\tablegroup}[1]{\textcolor{tableaccent}{\bfseries #1}}

\newcounter{resultstablebefore}
\usepackage{url}
\usepackage[breaklinks,hypertexnames=false]{hyperref}
\hypersetup{
  hidelinks,
  pdftitle={Revisiting Structure-Based Filtering for AI-Designed Antibodies: Opportunities, Limits, and What Matters},
  pdfauthor={Langzhang Liang, Juncheng Qian, Xiaojie Ma, Mingxi Zou, Yinghui Xu, Zenglin Xu, Tianlei Ying}
}

\title{Revisiting Structure-Based Filtering for AI-Designed Antibodies: Opportunities, Limits, and What Matters}

\author{%
\parbox{\dimexpr\textwidth-2\tabcolsep\relax}{\centering\normalfont
Langzhang Liang\textsuperscript{1,2}, Juncheng Qian\textsuperscript{2,3},
Xiaojie Ma\textsuperscript{3}, Mingxi Zou\textsuperscript{1},\\[0.3em]
Yinghui Xu\textsuperscript{1}, Zenglin Xu\textsuperscript{1,4,*},
Tianlei Ying\textsuperscript{2,3,*}\\[0.7em]
\textsuperscript{1}\,AI\textsuperscript{3} Institute, Fudan University\\
\textsuperscript{2}\,Genesius Lab, Shanghai Innovation Institute\\
\textsuperscript{3}\,School of Basic Medical Sciences, Fudan University\\
\textsuperscript{4}\,Shanghai Academy of AI for Science\\[0.5em]
\textsuperscript{*}\,Corresponding authors:
\href{mailto:zenglin@gmail.com}{\nolinkurl{zenglin@gmail.com}},
\href{mailto:tlying@fudan.edu.cn}{\nolinkurl{tlying@fudan.edu.cn}}\\[0.3em]
Langzhang Liang:
\href{mailto:lazylzliang@gmail.com}{\nolinkurl{lazylzliang@gmail.com}}
}}

\newcommand{\ind}{\mathbb{1}}

\theoremstyle{definition}
\newtheorem{definition}{Definition}

\iclrfinalcopy
\begin{document}
\raggedbottom
\maketitle

\begin{abstract}
AI antibody design can generate far more candidates than can be experimentally tested, making post-generation filtering a central bottleneck. We study whether structure-prediction outputs can help prioritize AI-designed antibodies for experimental validation. We evaluate this problem on 4,184 experimentally characterized antibody--antigen pairs, combining newly generated wet-lab measurements with curated affinity data across diverse antigens and assay settings. We introduce two complementary filtering signals: Regional Interaction Contrast (RIC), which captures support for the predicted antibody--antigen interface, and Non-CDR Antibody Confidence (NAC), which measures antibody confidence outside the CDRs. Across AlphaFold~3, Protenix-v2, ESMFold2, and OpenDDE, we find a striking difference in what information matters for filtering. Interface evidence is strongest when comparing candidates designed for different antigens, whereas antibody-level reliability is substantially more informative for ranking candidates within the same antigen. In particular, non-CDR antibody confidence improves the average within-antigen filtering rank over whole-antibody confidence across predictors, revealing an unexpected source of filtering signal beyond the predicted interface. Finally, we examine the information encoded in the predictor outputs themselves. Predicted structures contain substantial but incomplete binding-related signal, revealing both the opportunities and limits of structure-based filtering: better metrics can exploit current predictors more effectively, but their performance ultimately remains constrained by what those predictors can distinguish. Code is available in the \href{https://github.com/Jinx-byebye/Antibody-filtering}{code repository}.

\end{abstract}

\section{Introduction}
\label{sec:introduction}

Antibodies are highly versatile molecular-recognition proteins whose ability to
bind specific targets underlies their widespread use in therapeutics,
diagnostics, and biological research. Conventional discovery relies on
immunization, library screening, and affinity maturation
\citep{chao2006yeastdisplay}. Computational methods increasingly complement
these approaches by directly generating antigen-conditioned sequences,
backbones, or joint sequence--structure designs
\citep{luo2022diffab,kong2023dymean}. Recent diffusion-, flow-matching-, and
all-atom generative systems have further extended this paradigm to
experimentally validated epitope-specific antibodies, single-domain
antibodies, and other protein binders
\citep{bennett2026rfantibody,stark2025boltzgen,yu2026ppiflow}. As a result,
large and structurally diverse libraries of candidate binders can now be
generated entirely \emph{in silico}.

This progress does not eliminate the experimental bottleneck. A computational
campaign can readily produce hundreds or thousands of candidates, whereas
synthesis, expression, binding assays, and downstream characterization remain
comparatively costly and capacity limited. Deciding \emph{which generated
candidates should be tested} has therefore become a central post-generation
problem. Candidate selection often begins with sequence-level criteria, such
as removing developability liabilities, undesirable motifs, or poorly behaved
sequences. Such filters are valuable, but provide limited evidence about the
target-specific question that ultimately determines experimental success:
whether a candidate actually binds its intended antigen.

Protein-complex predictors provide a natural source of such target-specific
evidence. AlphaFold-Multimer and AlphaFold~3, together with more recent
structure-prediction models, predict biomolecular complexes while providing
coordinates together with residue-, pair-, and interface-level confidence or
interaction estimates
\citep{evans2022multimer,abramson2024alphafold3,
zhang2026protenix,candido2026esmfold2,passaro2025boltz2}. Scores derived from
these predictions, including ipTM, pDockQ, pDockQ2, and AntiConf, provide
readily available summaries of predicted complex or interface confidence
\citep{bryant2022pdockq,zhu2023pdockq2,unsal2026anticonf}. Structural
confidence and interface properties have consequently become important
prioritization signals in computational binder-design pipelines
\citep{pacesa2025bindcraft}.

For antibody discovery, however, the relevant decision is not simply whether a
predicted complex appears credible. An effective \emph{in silico} filter must
balance two competing errors. False positives consume limited experimental
capacity on candidates that do not bind, increasing the cost of discovery.
False negatives are also costly: successful designs may be rare for difficult
antigens, and an overly restrictive filter can discard the few true binders
produced by a generator. Recovering such hits may require another full cycle of
generation, prediction, and screening. The practical objective is therefore
to retain as many true binders as possible while keeping the false-positive
rate and experimental testing burden under control.

This decision problem raises a more fundamental question about the use of
structure prediction for antibody filtering: given the predicted structures
and confidence outputs for a candidate, how much information about binding is
present, and how effectively can that information be organized
into candidate-selection scores? We therefore study \textbf{structure-based
\emph{in silico} filtering of designed antibodies}: whether predictor outputs
can provide structural evidence that more accurately prioritizes
AI-designed antibody candidates by their likelihood of binding the intended
antigen. We evaluate this question on a large and heterogeneous collection of
experimentally characterized antibody--antigen pairs spanning diverse assay
modalities, evidence sources, and quantitative affinity measurements.

Our contributions are fourfold: \begin{itemize} \item \textbf{A practical formulation of structure-based antibody filtering.} We study post-generation filtering of AI-designed antibodies using only structure-prediction outputs, with evaluation centered on recovering true binders under realistic false-positive and experimental-budget constraints. \item \textbf{A diverse experimental benchmark.} We assemble 4,184 experimentally characterized antibody--antigen candidates spanning multiple antigens and assay modalities, combining curated literature data with newly generated wet-lab measurements. \item \textbf{Two complementary filtering metrics.} We introduce Regional Interaction Contrast (RIC) to capture interface-level support across predicted structures, and Non-CDR Antibody Confidence (NAC) to measure the reliability of the predicted antibody outside the CDRs. 

\item \textbf{An unexpected finding on antibody reliability.} 
Across four structure predictors, antibody-level reliability emerges as the strongest signal for within-antigen filtering, while interface evidence remains more effective when comparing candidates designed for different antigens. Surprisingly, confidence outside the CDRs improves average within-antigen filtering rank over whole-antibody confidence, revealing a robust filtering signal that is largely overlooked by existing interface-centric approaches.

\end{itemize}

\section{Background and Problem Formulation}
\label{sec:problem}

\subsection{AI Antibody Design as a Generate--Predict--Filter Pipeline}
\label{sec:pipeline}

\paragraph{From generation to filtering.}
An AI antibody design workflow begins with an antigen and constraints such as
the antibody format, framework scaffold, CDR lengths, residues that may be
redesigned, and a desired epitope or set of target hotspots. \textbf{Generate.} Current systems
span several generative paradigms. RFantibody first generates a docked antibody
backbone and then assigns sequences with ProteinMPNN
\citep{bennett2026rfantibody}; PPIFlow couples backbone flow matching with
sequence design and in silico maturation \citep{yu2026ppiflow}; and Chai-2
directly supports zero-shot antibody design \citep{chai22025}. More general
all-atom systems, including BoltzGen and ODesign, jointly reason over binder and
target structure and can generate antibody-like or other binder modalities
\citep{stark2025boltzgen,zhang2025odesign}. Despite these architectural
differences, generation is only the first stage. \textbf{Predict.} Candidate sequences are
screened for sequence validity and developability, and their antibody--antigen
complex structures are independently predicted. \textbf{Filter.} Designs are then assessed
using the predicted complexes and physicochemical quantities such as interface
confidence, buried surface area, hydrogen bonding, shape complementarity, and
Rosetta interface-energy estimates. The final experimental set is chosen under a
synthesis budget by combining predicted complex quality, interface quality,
developability, and diversity.

\paragraph{In silico filtering with structural confidence.}
The central step in post-design filtering is to predict each designed
antibody--antigen complex with a co-folding model. Alongside coordinates,
co-folding models report confidence metrics used directly for screening:
pLDDT measures local residue-level confidence, ipTM measures confidence in the
relative placement of chains, PAE measures positional uncertainty after
alignment, and PDE or gPDE summarizes uncertainty in predicted inter-residue
distances. These metrics are attractive because they directly probe whether
the antibody is well formed and placed against the antigen through a confident
interface. Antibody--antigen prediction remains
challenging, however, because these complexes often lack strong paired
evolutionary information and concentrate binding determinants in flexible,
hypervariable CDR loops. Errors in CDR geometry or docking therefore propagate
directly into filtering and reduce downstream hit recovery
\citep{unsal2026anticonf,aintibody2026benchmark}. How best to use predicted
complexes and their confidence outputs for in silico screening therefore
remains an open question.

\subsection{Quantifying the Utility of Structural Filters}
\label{sec:objective}

To compare structural filters, we need to quantify how effectively they turn
predicted complex confidence into useful experimental candidates. In silico
filtering sits between a scalable computational pipeline and a much more
constrained experimental stage. Two practical facts determine what makes a
filter useful:
\par\noindent\textbf{Experimental validation is expensive and capacity limited.}
Sequence generation and structure prediction can be parallelized over many
candidates, whereas synthesis, expression, and functional testing remain
limited by cost, time, and laboratory capacity. False positives therefore
spend scarce experimental resources on candidates that passed the
computational screen but fail in the laboratory.

\noindent\textbf{Successful computational designs remain rare.}
Current antibody designers have limited and strongly target-dependent hit
rates. A low-recall filter can reject the few genuine successes before they
are tested, forcing additional cycles of generation, prediction, and
screening.

An effective filter must therefore suppress false positives without doing so
by discarding most true binders.

\paragraph{Filtering decision.}
Consider a designed antibody $a$ in design context $c$, which includes its
antigen target and design constraints. Let $M_{ac}$ be a structural-confidence
score such as ipTM. Given a prespecified threshold $\sigma$, the filter retains
the candidate for subsequent analysis or experimental testing when
\begin{equation}
\widehat A_{ac}(\sigma)=\ind\!\left\{M_{ac}\geq\sigma\right\}.
\label{eq:filtering_decision}
\end{equation}
Here $\widehat A_{ac}=1$ means that the candidate advances. In practice, $M_{ac}$ may be a single metric or a composite score combining multiple structural criteria, such as interface confidence, antibody pLDDT, and inter-chain error. In this paper, we focus on individual metrics.

\begin{definition}[Filtering outcomes]
$\mathcal D=\{(a,c,Y_{ac})\}$ denotes a set of designed antibodies, their
design contexts, and their \textbf{wet-lab} validation outcomes
$Y_{ac}\in\{0,1\}$. Here $Y_{ac}=1$ means that candidate $a$ satisfies the
prespecified experimental binding criterion in context $c$ and is treated as
a binder; $Y_{ac}=0$ denotes a negative outcome. A filter advances the
candidate to synthesis or wet-lab validation when $\widehat A_{ac}=1$ ($M_{ac}\geq\sigma$).
Comparing $\widehat A_{ac}$ with $Y_{ac}$ yields true positives (TP), false
positives (FP), true negatives (TN), and false negatives (FN). The precision
$P$, recall $R$, false-positive rate $F$, and selected fraction $Q$ are
\begin{equation}
P=\frac{\mathrm{TP}}{\mathrm{TP}+\mathrm{FP}},
\quad
R=\frac{\mathrm{TP}}{\mathrm{TP}+\mathrm{FN}},
\quad
F=\frac{\mathrm{FP}}{\mathrm{FP}+\mathrm{TN}},
\quad
Q=\frac{\mathrm{TP}+\mathrm{FP}}{|\mathcal D|}.
\label{eq:filtering_outcomes}
\end{equation}
\end{definition}

\begin{definition}[Filtering utility]
Let $\underline R$ be the 95\% Wilson lower confidence bound for recall, and
let $\overline F$ be the corresponding upper confidence bound for the
false-positive rate. Let $f_{\mathrm{safe}}$
be an acceptable FPR and $f_{\mathrm{severe}}>f_{\mathrm{safe}}$ mark the
onset of severe FPR penalty. Define
\begin{equation}
W(F)=\exp\!\left[-\left(\frac{F}{f_{\mathrm{severe}}}\right)^2
-\left(\frac{(F-f_{\mathrm{safe}})_+}
{f_{\mathrm{severe}}-f_{\mathrm{safe}}}\right)^2\right],
\qquad (x)_+=\max(x,0).
\label{eq:utility_gates}
\end{equation}
For fixed $f_{\mathrm{safe}}$ and $f_{\mathrm{severe}}$, the utility
at threshold $\sigma$ and the best utility of metric $m$ are
\begin{equation}
U(m,\sigma;\mathcal D)=\underline R\,W(\overline F),\qquad
U^\star(m;\mathcal D)=\max_{\sigma\in\mathbb R}U(m,\sigma;\mathcal D).
\label{eq:filtering_utility}
\end{equation}
\end{definition}

\paragraph{Interpreting the utility.}
The first term in $W$ gradually penalizes FPR even below
$f_{\mathrm{safe}}$, favoring cleaner filters when recall is similar. Above
$f_{\mathrm{safe}}$, the second term steepens the penalty: gains in recall
must then offset a greater cost from false positives. The penalty remains
smooth, with no hard cutoff. We set $f_{\mathrm{safe}}=0.10$ and
$f_{\mathrm{severe}}=0.30$; $W$ equals $0.97$, $0.89$, $0.50$, and
$e^{-2}\approx0.14$ at FPRs of 5\%, 10\%, 20\%, and 30\%, respectively.
Using the Wilson lower bound for recall and upper bound for FPR also accounts
for uncertainty in the observed rates.

\section{Method}
\label{sec:method}
In this section, we introduce two complementary structure-based filtering
metrics, RIC and NAC, which capture interface-level and antibody-level evidence
from predicted complex ensembles, respectively.

\subsection{Candidate Evidence}
\label{sec:candidate_evidence}

For each antibody--antigen candidate $i$, let $X_i$ be the ensemble of
complexes predicted across independent runs (called seeds in the inference scripts) and their samples.
These predictions provide two kinds of evidence worth examining. First, do
the antigen regions predicted to contact the antibody CDRs receive stronger,
reproducible support than the next-ranked antigen regions? Second, how confidently
is the antibody itself modeled outside the more variable CDR loops? We capture
these questions with Regional Interaction Contrast (RIC) and Non-CDR Antibody
Confidence (NAC). We represent each candidate by these two complementary sources of evidence:
\[
\mathbf E_i=(\operatorname{RIC}(X_i),\operatorname{NAC}(X_i)).
\]
For simplicity, we omit the candidate index $i$ below. We denote $\mathcal A$,
$\mathcal C$, and $\mathcal G$ as the residue sets of the antibody,
CDR regions, and antigen, respectively.

\subsection{Regional Interaction Contrast}
\label{sec:ric}

High confidence at an entire interface need not single out the CDR contacts
predicted for a candidate. RIC instead compares the most strongly predicted
antigen contacts of each CDR residue with the next-ranked antigen region. It
asks whether the first region also has stronger support from predicted aligned
error (PAE) across two different runs. Comparing regions reduces the influence
of uniformly high agreement, while comparing runs tests whether the contact
pattern persists across predictions.
The next-ranked contacts provide plausible alternatives to
the proposed contact region. A reference run defines both regions; PAE from
that run and another run tests their relative support.

For a reference run $r$, let $\bar C_r(a,b)$ be the mean bidirectional contact
probability over its samples for CDR residue $a$ and antigen residue $b$, and
let $h_r(a)=\max_{b\in\mathcal G}\bar C_r(a,b)$. Ranking antigen residues by
$\bar C_r(a,b)$ defines normalized weights $w^H_{ra}(b)$ and $w^L_{ra}(b)$ for
the highest- and next-ranked regions. Let $A_{rtjk}(a,b)$ measure bidirectional
PAE agreement between sample $j$ in run $r$ and sample $k$ in another run $t$;
larger PAE lowers agreement.
At $\tau=16$~\AA, a directional PAE of $8$~\AA\ in each run gives agreement $1/2$.
Write $\Delta_{rtjk}(a)$ for the difference in PAE agreement
between the two regions of CDR residue $a$, and $z_{rtjk}$ for the score of
that sample pair. The candidate score is their nested median:
\begin{equation}
\begin{gathered}
\Delta_{rtjk}(a)
=\sum_{b\in\mathcal G}\bigl[w^H_{ra}(b)-w^L_{ra}(b)\bigr]A_{rtjk}(a,b),
\qquad
z_{rtjk}
=\frac{\sum_{a\in\mathcal C}h_r(a)^2\Delta_{rtjk}(a)}
{\sum_{a\in\mathcal C}h_r(a)},\\[-2pt]
\operatorname{RIC}(X)
=\operatorname*{median}_{r}\operatorname*{median}_{t\ne r}
\operatorname*{median}_{j,k}z_{rtjk}.
\end{gathered}
\label{eq:ric_score}
\end{equation}
The squared contact support emphasizes CDR residues with stronger predicted contacts;
the nested medians require the contrast to persist across sample pairs and
runs. The contact average, PAE agreement, rank weights, and zero-support rule
are specified in Appendix~\ref{app:ric_implementation}.

\subsection{Non-CDR Antibody Confidence}
\label{sec:nac}

NAC measures confidence in the antibody outside the CDRs. CDR loops are
structurally variable, so excluding them gives a broader view of how
confidently the rest of the antibody is modeled within the complex. For each
sample, we average pLDDT over the non-CDR residues, then take the median over
samples within a run and the median across runs. With $p_{rj}^{\mathrm{nc}}$
denoting the non-CDR confidence of sample $j$ in run $r$, this gives
\begin{equation}
\operatorname{NAC}(X)
=\operatorname*{median}_{r}\operatorname*{median}_{j\in\mathcal K_r}
p_{rj}^{\mathrm{nc}},\qquad
p_{rj}^{\mathrm{nc}}=
\frac{\sum_{a\in\mathcal A\setminus\mathcal C}
\operatorname{pLDDT}_{rja}}{100|\mathcal A\setminus\mathcal C|}.
\label{eq:nac_score}
\end{equation}
NAC is computed from the predicted complex, so its confidence values also
reflect the context in which the antibody and antigen were modeled.

\section{Experiments}
\label{sec:experiments}

\subsection{Setup}
\label{sec:setup}

\paragraph{Public benchmarks.}
We curate two benchmarks from public sources. IgDesign is an experimentally
validated inverse-folding method that redesigns heavy-chain CDRs on a native
antibody--antigen backbone and measures binding by surface plasmon resonance
(SPR)~\citep{shanehsazzadeh2023igdesign}. We also separately
curate antibody--antigen pairs with direct affinity measurements reported in
papers and patents. 

\paragraph{Wet-lab validation.}
We additionally evaluate display-selected antibodies against
seven antigens. TfR, spike S2, and GPC3 were screened by yeast display; CD98hc,
EGFR, PLXNA1 ectodomain, and TNFRSF23 by phage display. Sequencing across
selection rounds tracks clone enrichment. For each antigen, after sequence
quality control and deduplication, we
classify 100 late-enriched candidates as putative binders and
100 baseline or early low-enrichment candidates as putative non-binders.
Table~\ref{tab:datasets}
summarizes the three benchmarks; Appendix~\ref{app:benchmark_construction}
describes their construction, eligibility criteria, and filtering in detail.

\paragraph{Co-folding model training-set exposure.}
AlphaFold~3, Protenix-v2, and ESMFold2 each use a 30 September 2021 cutoff
for PDB training structures~\citep{candido2026esmfold2}. ESMFold2 also trains
on AlphaFold DB-predicted monomers. An exact-match audit against SAbDab found
no pre-cutoff structure matching both the antibody variable chain(s) and
antigen construct among the 4,184 candidates; thus, no exact complex was
identified in the models' PDB structural training data.
This audit covers the three predictors named above, not OpenDDE's training data.

\paragraph{Structure prediction protocol.}
For each of the 4,184 antibody--antigen pairs, we independently predict the
complex with AlphaFold~3, Protenix-v2, ESMFold2, and OpenDDE~\citep{opendde2026}. Every predictor uses
MSAs, no structural templates, 10 random seeds, and 5 samples per seed, yielding
up to 50 predicted complexes per pair and predictor. All subsequent metrics
are computed from these predicted structures and their associated confidence
outputs. Predictor configurations and inference details are given in
Appendix~\ref{app:experimental_details}.

\paragraph{Baselines.}
Baselines include predictor-native ranking confidence, pair-iPTM, antibody and
interface pLDDT, directional inter-chain PAE, and PDE/gPDE. We also evaluate
pDockQ~\citep{bryant2022pdockq}, pDockQ2~\citep{zhu2023pdockq2}, three
variants of the interface score derived from predicted aligned error
(ipSAE)~\citep{dunbrack2025ipsae}, and AntiConf~\citep{unsal2026anticonf}.
Further baseline implementation details are provided in
Appendix~\ref{app:baseline_details}.

\paragraph{Threshold selection.}
Each metric advances a candidate when its score satisfies
$M\geq\sigma$, where $\sigma$ is a threshold. To compare metrics on equal terms, we sweep all attainable
thresholds and select the value that maximizes the filtering utility in
Definition~2. Portfolio-wide (\textit{global}) filtering uses one
threshold shared across all antigens, whereas campaign-specific
(\textit{target-specific}) filtering optimizes the decision rule separately within
each antigen. We report the resulting
utility and precision, recall at FPR levels of 10\%, 20\%, and 30\%, and recall
when the selected fraction is fixed at 5\% or 20\%. The last view directly
models a limited synthesis budget: candidates are ranked by the metric and
only the highest-scoring fraction is tested. Further details on
threshold selection are given in Appendix~\ref{app:threshold_selection}.

\begin{table*}[!htbp]
\centering
\small
\setlength{\tabcolsep}{2.5pt}
\renewcommand{\arraystretch}{0.94}
\caption{\textbf{Benchmark dataset statistics.} Unique-antigen counts
refer to distinct antigen constructs within each benchmark.}
\label{tab:datasets}
\begin{tabularx}{\textwidth}{@{}>{\raggedright\arraybackslash}p{3.5cm}rrrrr>{\raggedright\arraybackslash}X@{}}
\tabletoprule
Dataset & Unique antigens & Positive & Negative & Total & Positive rate & Label source \\
\tablemidrule
Display-based enrichment & 7 & 700 & 700 & 1,400 & 50.0\% & round-wise selection enrichment \\
IgDesign SPR & 7 & 249 & 922 & 1,171 & 21.3\% & replicated SPR binding \\
Literature/patent affinity & 165 & 1,510 & 103 & 1,613 & 93.6\% & direct SPR/BLI/ITC $K_D$ \\
\tablemidrule
Total & 179 & 2,459 & 1,725 & 4,184 & 58.8\% & \\
\tablebottomrule
\end{tabularx}
\end{table*}

\subsection{Results}
\label{sec:results}

Tables~\ref{tab:global_results}
and~\ref{tab:target_specific_results} evaluate structure-based filtering from
two practical perspectives. The first examines portfolio-level selection,
where candidates designed for different antigens compete under a shared
threshold, while the second examines within-antigen ranking under
target-specific thresholds. The main observations from these complementary
evaluations are summarized below.

\captionof{table}{\textbf{In silico filtering using a single global threshold shared across antigen targets.} Recall and precision at $U^\star$ use the threshold that maximizes filtering utility. Recall and precision are expressed as percentages. The benchmark includes 4,184 candidates per predictor, comprising 2,459 positives and 1,725 negatives. Best values are bold and second-best values are underlined. For ESMFold2, 275 out-of-memory complexes are excluded. \textsuperscript{*}gRIC is used for ESMFold2 because the model does not provide predicted contact probabilities; it approximates RIC using 8-\AA{} geometric contacts from predicted coordinates.}
\label{tab:global_results}
\setcounter{resultstablebefore}{\value{table}}
\begingroup
\scriptsize
\setlength{\tabcolsep}{0.9pt}
\renewcommand{\arraystretch}{0.96}
\begin{xltabular}{\textwidth}{@{}p{1.75cm}*{9}{>{\centering\arraybackslash}X}@{}}
\tabletoprule
\textbf{Metric} &
\shortstack{\textbf{Recall @}\\\textbf{10\% FPR}} &
\shortstack{\textbf{Recall @}\\\textbf{20\% FPR}} &
\shortstack{\textbf{Recall @}\\\textbf{30\% FPR}} &
\shortstack{\textbf{Recall @}\\$\boldsymbol{Q=5\%}$} &
\shortstack{\textbf{Recall @}\\$\boldsymbol{Q=20\%}$} &
\shortstack{\textbf{Utility}\\$\boldsymbol{U^\star}$} &
\shortstack{\textbf{Recall}\\\textbf{@ $U^\star$}} &
\shortstack{\textbf{Precision}\\\textbf{@ $U^\star$}} &
\shortstack{\textbf{Average}\\\textbf{rank}} \\
\tablemidrule
\endfirsthead
\multicolumn{10}{@{}l@{}}{\textit{Table~\ref{tab:global_results} continued}}\\
\tabletoprule
\textbf{Metric} &
\shortstack{\textbf{Recall @}\\\textbf{10\% FPR}} &
\shortstack{\textbf{Recall @}\\\textbf{20\% FPR}} &
\shortstack{\textbf{Recall @}\\\textbf{30\% FPR}} &
\shortstack{\textbf{Recall @}\\$\boldsymbol{Q=5\%}$} &
\shortstack{\textbf{Recall @}\\$\boldsymbol{Q=20\%}$} &
\shortstack{\textbf{Utility}\\$\boldsymbol{U^\star}$} &
\shortstack{\textbf{Recall}\\\textbf{@ $U^\star$}} &
\shortstack{\textbf{Precision}\\\textbf{@ $U^\star$}} &
\shortstack{\textbf{Average}\\\textbf{rank}} \\
\tablemidrule
\endhead
\multicolumn{10}{@{}r@{}}{\textit{Continued on next page}}\\
\endfoot
\endlastfoot
\multicolumn{10}{@{}l@{}}{\tablegroup{AlphaFold 3}} \\
Pair-iPTM & 14.64 & 19.03 & \textbf{31.35} & \underline{8.48} & 19.34 & 0.1150 & 14.84 & 67.47 & \underline{3.12} \\
Interface PAE $\downarrow$ & 9.88 & 15.45 & 26.68 & 5.21 & 17.89 & 0.0815 & 13.58 & 56.71 & 8.12 \\
Antibody pLDDT & 7.40 & 16.10 & 18.71 & 5.53 & 17.32 & 0.0794 & 13.42 & 56.12 & 9.25 \\
Ranking score & \underline{14.92} & 16.63 & 25.82 & \textbf{8.50} & 18.58 & \underline{0.1229} & 14.92 & \underline{72.53} & 3.62 \\
pDockQ & 10.33 & 14.03 & 30.42 & 6.67 & 17.97 & 0.0791 & 10.33 & 59.76 & 7.44 \\
pDockQ2 & 12.48 & 15.41 & 26.88 & 5.12 & 18.06 & 0.0972 & 12.77 & 63.56 & 6.94 \\
ipSAE-d0res & 9.72 & 16.31 & 27.45 & 5.33 & 18.06 & 0.0831 & 13.62 & 57.17 & 6.94 \\
ipSAE-d0dom & 8.74 & 16.92 & 27.17 & 4.96 & 17.85 & 0.0856 & \underline{15.21} & 57.98 & 6.75 \\
ipSAE-d0chn & 14.07 & \underline{19.28} & \underline{30.46} & \textbf{8.50} & \underline{19.40} & 0.1119 & 13.91 & 69.09 & \underline{3.12} \\
AntiConf & 2.81 & 6.99 & 19.15 & 2.56 & 15.86 & 0.0284 & 13.95 & 44.66 & 10.88 \\
\textbf{RIC (Interface)} & \textbf{17.93} & \textbf{21.43} & 28.95 & \textbf{8.50} & \textbf{21.27} & \textbf{0.1465} & \textbf{17.61} & \textbf{75.83} & \textbf{1.50} \\
\textbf{NAC (Antibody)} & 9.80 & 14.52 & 17.61 & 4.80 & 16.27 & 0.0767 & 10.33 & 57.86 & 10.31 \\
\tablemidrule
\multicolumn{10}{@{}l@{}}{\tablegroup{Protenix-v2}} \\
Pair-iPTM & \underline{11.43} & 20.86 & 29.89 & \underline{7.97} & 20.33 & \underline{0.0904} & 11.14 & \underline{66.67} & \underline{3.00} \\
Pair-gPDE $\downarrow$ & 2.89 & 5.57 & 13.79 & 2.72 & 13.42 & 0.0223 & 3.38 & 29.96 & 12.56 \\
Interface PAE $\downarrow$ & 9.92 & 15.90 & 27.12 & 4.60 & 18.02 & 0.0824 & 12.24 & 57.44 & 5.81 \\
Antibody pLDDT & 3.90 & 10.90 & 16.96 & 3.13 & 14.84 & 0.0432 & 9.39 & 43.10 & 9.25 \\
Ranking score & 3.13 & 8.38 & 27.12 & 2.81 & 17.16 & 0.0415 & 15.62 & 49.42 & 8.69 \\
pDockQ & 6.71 & 9.48 & 21.27 & 4.96 & 15.13 & 0.0514 & 7.08 & 48.74 & 8.25 \\
pDockQ2 & 9.96 & 15.94 & 27.00 & 4.27 & 17.97 & 0.0852 & 12.08 & 58.81 & 5.88 \\
ipSAE-d0res & 6.75 & 16.84 & 29.16 & 5.57 & 18.83 & 0.0639 & 16.63 & 54.53 & 5.25 \\
ipSAE-d0dom & 6.67 & 18.10 & 29.20 & 4.92 & 19.36 & 0.0685 & \underline{18.42} & 56.62 & 5.00 \\
ipSAE-d0chn & 11.39 & \underline{22.12} & \underline{32.98} & 7.89 & \underline{20.37} & 0.0892 & 11.02 & 66.10 & 3.25 \\
AntiConf & 3.21 & 5.49 & 16.51 & 2.93 & 14.07 & 0.0227 & 3.17 & 34.21 & 11.62 \\
\textbf{RIC (Interface)} & \textbf{27.90} & \textbf{32.53} & \textbf{33.75} & \textbf{8.46} & \textbf{27.65} & \textbf{0.2309} & \textbf{27.49} & \textbf{82.14} & \textbf{1.00} \\
\textbf{NAC (Antibody)} & 3.50 & 8.09 & 14.68 & 2.52 & 13.42 & 0.0321 & 6.71 & 36.03 & 11.44 \\
\tablemidrule
\multicolumn{10}{@{}l@{}}{\tablegroup{ESMFold2}} \\
Pair-iPTM & 1.57 & 3.80 & 11.05 & 1.48 & 11.63 & 0.0132 & 4.16 & 21.38 & 10.62 \\
Interface PAE $\downarrow$ & 3.18 & 10.20 & 15.07 & 2.95 & 14.18 & 0.0379 & 8.77 & 39.20 & 7.25 \\
Antibody pLDDT & 3.40 & 7.29 & 11.85 & 2.82 & 12.21 & 0.0314 & 6.26 & 33.98 & 8.31 \\
pDockQ & \underline{8.09} & 14.18 & 20.93 & \underline{5.28} & 17.35 & 0.0646 & 9.79 & 50.34 & 4.38 \\
pDockQ2 & 3.62 & 11.58 & 18.56 & 2.86 & 16.23 & 0.0457 & 14.40 & 47.49 & 6.00 \\
ipSAE-d0res & 4.87 & 22.00 & 27.59 & 3.67 & 20.44 & 0.0833 & \textbf{22.00} & 59.56 & 3.00 \\
ipSAE-d0dom & 4.79 & 20.71 & 27.46 & 3.35 & 20.21 & 0.0799 & 19.86 & 57.81 & 4.25 \\
ipSAE-d0chn & 7.69 & \underline{24.46} & \underline{31.98} & 5.14 & \underline{21.33} & \underline{0.0984} & \underline{21.78} & \underline{61.41} & \underline{2.25} \\
AntiConf & 1.79 & 7.29 & 16.32 & 1.79 & 14.40 & 0.0258 & 7.78 & 33.59 & 8.56 \\
\textbf{gRIC$^{*}$} & \textbf{23.03} & \textbf{28.58} & \textbf{33.68} & \textbf{8.72} & \textbf{25.36} & \textbf{0.1868} & 21.69 & \textbf{81.38} & \textbf{1.25} \\
\textbf{NAC (Antibody)} & 2.42 & 4.65 & 7.74 & 2.24 & 11.00 & 0.0205 & 3.80 & 25.68 & 10.12 \\
\tablemidrule
\multicolumn{10}{@{}l@{}}{\tablegroup{OpenDDE}} \\
Pair-iPTM & 12.97 & 29.93 & 40.18 & 6.34 & \underline{23.14} & 0.1375 & 24.32 & 68.50 & 4.38 \\
Pair-gPDE $\downarrow$ & 4.51 & 28.06 & 37.82 & 3.46 & 21.55 & 0.1142 & 25.58 & 66.28 & 7.88 \\
Interface PAE $\downarrow$ & 13.42 & 28.95 & 37.54 & 7.04 & 21.96 & 0.1129 & 26.64 & 66.57 & 5.88 \\
Antibody pLDDT & 12.32 & 20.09 & 30.22 & 6.59 & 20.01 & 0.0965 & 12.20 & 65.36 & 9.50 \\
Ranking score & 3.50 & 26.39 & 33.63 & 2.36 & 21.47 & 0.1057 & 23.10 & 64.33 & 10.00 \\
pDockQ & 7.12 & 23.18 & 37.90 & 5.04 & 20.82 & 0.0961 & 15.58 & 60.51 & 9.50 \\
pDockQ2 & 13.99 & 28.91 & 38.51 & 6.63 & 22.16 & 0.1189 & \textbf{27.82} & 67.59 & 4.62 \\
ipSAE-d0res & 6.71 & 30.26 & 40.42 & 3.33 & 22.69 & 0.1268 & 22.61 & 66.83 & 6.38 \\
ipSAE-d0dom & 8.46 & 30.22 & \underline{42.98} & 3.42 & 22.77 & 0.1287 & 22.53 & 67.07 & 5.75 \\
ipSAE-d0chn & \underline{14.31} & \underline{31.52} & \textbf{46.89} & \underline{7.16} & 23.10 & \underline{0.1396} & 24.56 & \underline{68.79} & \underline{2.38} \\
AntiConf & 2.81 & 19.52 & 34.73 & 2.28 & 19.89 & 0.0831 & 24.40 & 62.37 & 11.12 \\
\textbf{RIC (Interface)} & \textbf{27.12} & \textbf{34.61} & 39.77 & \textbf{7.73} & \textbf{27.12} & \textbf{0.2196} & \underline{27.12} & \textbf{79.78} & \textbf{1.62} \\
\textbf{NAC (Antibody)} & 6.26 & 11.02 & 21.27 & 4.51 & 15.82 & 0.0493 & 8.82 & 45.02 & 12.00 \\
\tablebottomrule
\end{xltabular}
\endgroup
\setcounter{table}{\value{resultstablebefore}}
\vspace{4pt}
{%
\begingroup
\setlength{\fboxsep}{5pt}
\noindent\fcolorbox{tableline}{gray!4}{%
\begin{minipage}{\dimexpr\linewidth-2\fboxsep-2\fboxrule\relax}
\footnotesize
\setlength{\smallskipamount}{1pt}
\textbf{Key Messages from Tables~\ref{tab:global_results}
and~\ref{tab:target_specific_results}.}

\smallskip
\textbf{Interface evidence leads in across-antigen filtering.}
With one threshold shared across antigens in
Table~\ref{tab:global_results},
RIC has the best average rank for
AlphaFold~3 (1.50) and Protenix-v2 (1.00), with large gains in low-FPR recall and
utility. Its geometric-contact approximation, gRIC, also ranks first for
ESMFold2 (1.25), whereas NAC ranks near the bottom (10.12).
OpenDDE shows the same pattern: RIC ranks first (1.62), while NAC ranks 12.00.

\smallskip
\textbf{Antibody reliability is strongest within an antigen.}
The ordering reverses when candidates compete only within the same antigen.
NAC ranks first for AlphaFold~3 and ESMFold2 (1.38 for both) and remains among
the strongest scores for Protenix-v2 (2.12). In contrast, RIC falls to average
ranks of 10.25 and 13.00 for AlphaFold~3 and Protenix-v2, and gRIC ranks 10.62
for ESMFold2.
For OpenDDE, NAC ranks first (2.00), ahead of whole-antibody pLDDT (2.38),
whereas RIC ranks last (13.00).

\smallskip
\textbf{Confidence outside the CDRs is more informative than whole-antibody
confidence.}
In campaign selection, NAC outperforms antibody pLDDT in 28 of 32
predictor--measure comparisons: all eight for AlphaFold~3 and ESMFold2, seven
for Protenix-v2, and five for OpenDDE. Removing the CDRs therefore usually
improves filtering, despite discarding the residues most directly involved
in binding. OpenDDE illustrates the remaining trade-off: whole-antibody
pLDDT gives higher utility (0.1404 versus 0.1336), while NAC gives higher
recall at $Q=20\%$ (32.65\% versus 32.22\%).

\smallskip
\textbf{Predictor comparisons depend on the evaluation context.}
Comparing AlphaFold~3 with Protenix-v2 on ten shared baselines, AlphaFold~3 has higher values in 54 of 80
baseline--measure comparisons in across-antigen filtering
(Table~\ref{tab:global_results}), whereas Protenix-v2 leads in 77 of 80 comparisons
under campaign-specific evaluation (Table~\ref{tab:target_specific_results}).

\smallskip
\textbf{Within-antigen filtering offers greater fixed-budget gains.}
At a 20\% testing budget, most scores in Table~\ref{tab:global_results}
fall below the roughly 20\% recall expected from random selection.
In Table~\ref{tab:target_specific_results}, NAC reaches 35.82\%, 36.21\%,
39.22\%, and 32.65\% recall with AlphaFold~3, Protenix-v2, ESMFold2, and OpenDDE.
Within-antigen filtering therefore appears more effective for the predictors tested.
\end{minipage}}
\endgroup
}
\par

\paragraph{Reading precision and recall.}
In the full 4,184-candidate benchmark, 58.8\% are binders, so testing at $Q=20\%$
caps recall at 34.0\% even with perfect ranking. At $U^\star$,
a large fraction of selected candidates comes from the mostly positive
literature/patent benchmark. Such concentration can yield high precision while
leaving many binders elsewhere unselected. RIC/gRIC achieve FPRs of 8.0\%,
8.5\%, and 6.6\% for AlphaFold~3, Protenix-v2, and ESMFold2, respectively,
at $U^\star$.

\paragraph{Interpreting the contrasting signals.}
The recall gap between across-antigen and within-antigen filtering may arise from
target-dependent prediction errors. Candidates for one antigen can share
docking and confidence biases, which affect their scores similarly and partly
cancel in within-antigen ranking. Across antigens, different biases can shift
score scales, weakening a shared threshold. On the same candidates as
Table~\ref{tab:target_specific_results}, a shared global ranking still favors
RIC/gRIC over NAC (Table~\ref{tab:matched_population_portfolio}), although
its fixed-budget recall remains below random selection.
Figure~\ref{fig:regional_confidence_comparison}
also helps explain the strength of NAC: non-CDR pLDDT has the highest mean
within-antigen AUC in seven of eight predictor--benchmark comparisons. In six, AUC
decreases from non-CDR to whole-antibody, CDR, and interface pLDDT. This
ordering suggests a trade-off between binding relevance and prediction
reliability: confidence in the more reliably modeled non-CDR region can filter
better than confidence in the binding-proximal but harder-to-model CDRs or
predicted interface.

\captionof{table}{\textbf{Campaign-specific evaluation of 2,400 candidates (944 binders,
1,456 non-binders) across 13 display-based and IgDesign antigens.}
Candidates compete only within an antigen; literature/patent antigens have too
few negatives for low-FPR evaluation and are excluded. Appendix~\ref{app:campaign_specific_evaluation}
details eligibility, weighting, and results aggregation. For ESMFold2, 107
out-of-memory complexes are excluded.}
\label{tab:target_specific_results}
\setcounter{resultstablebefore}{\value{table}}
\begingroup
\scriptsize
\setlength{\tabcolsep}{0.9pt}
\renewcommand{\arraystretch}{0.94}
\begin{xltabular}{\textwidth}{@{}p{1.75cm}*{9}{>{\centering\arraybackslash}X}@{}}
\tabletoprule
\textbf{Metric} &
\shortstack{\textbf{Recall @}\\\textbf{10\% FPR}} &
\shortstack{\textbf{Recall @}\\\textbf{20\% FPR}} &
\shortstack{\textbf{Recall @}\\\textbf{30\% FPR}} &
\shortstack{\textbf{Recall @}\\$\boldsymbol{Q=5\%}$} &
\shortstack{\textbf{Recall @}\\$\boldsymbol{Q=20\%}$} &
\shortstack{\textbf{Utility}\\$\boldsymbol{U^\star}$} &
\shortstack{\textbf{Recall}\\\textbf{@ $U^\star$}} &
\shortstack{\textbf{Precision}\\\textbf{@ $U^\star$}} &
\shortstack{\textbf{Average}\\\textbf{rank}} \\
\tablemidrule
\endfirsthead
\multicolumn{10}{@{}l@{}}{\textit{Table~\ref{tab:target_specific_results} continued}}\\
\tabletoprule
\textbf{Metric} &
\shortstack{\textbf{Recall @}\\\textbf{10\% FPR}} &
\shortstack{\textbf{Recall @}\\\textbf{20\% FPR}} &
\shortstack{\textbf{Recall @}\\\textbf{30\% FPR}} &
\shortstack{\textbf{Recall @}\\$\boldsymbol{Q=5\%}$} &
\shortstack{\textbf{Recall @}\\$\boldsymbol{Q=20\%}$} &
\shortstack{\textbf{Utility}\\$\boldsymbol{U^\star}$} &
\shortstack{\textbf{Recall}\\\textbf{@ $U^\star$}} &
\shortstack{\textbf{Precision}\\\textbf{@ $U^\star$}} &
\shortstack{\textbf{Average}\\\textbf{rank}} \\
\tablemidrule
\endhead
\multicolumn{10}{@{}r@{}}{\textit{Continued on next page}}\\
\endfoot
\endlastfoot
\multicolumn{10}{@{}l@{}}{\tablegroup{AlphaFold 3}} \\
Pair-iPTM & 15.44 & 25.99 & 38.74 & 7.49 & 25.57 & 0.0807 & 20.68 & 51.54 & 8.00 \\
Interface PAE $\downarrow$ & 17.30 & 28.12 & 41.88 & 6.23 & 25.12 & 0.0847 & 22.51 & 46.82 & 7.88 \\
Antibody pLDDT & \underline{27.85} & \underline{42.27} & \underline{51.00} & \underline{8.91} & \underline{33.39} & \underline{0.1367} & 34.12 & \underline{60.36} & \underline{2.38} \\
Ranking score & 14.83 & 29.24 & 36.43 & 6.82 & 28.05 & 0.0870 & 23.48 & 49.33 & 6.25 \\
pDockQ & 19.63 & 33.50 & 44.64 & 6.60 & 26.00 & 0.1021 & 25.97 & 51.36 & 5.25 \\
pDockQ2 & 16.51 & 29.20 & 41.04 & 5.99 & 26.69 & 0.0869 & 23.42 & 48.75 & 6.75 \\
ipSAE-d0res & 12.01 & 17.69 & 21.78 & 7.58 & 26.64 & 0.0557 & \textbf{47.41} & 58.59 & 7.12 \\
ipSAE-d0dom & 10.62 & 17.69 & 22.54 & 8.52 & 26.32 & 0.0538 & 45.55 & 58.40 & 7.62 \\
ipSAE-d0chn & 10.61 & 17.69 & 22.54 & 8.52 & 26.02 & 0.0539 & \underline{45.85} & 57.53 & 7.75 \\
AntiConf & 17.31 & 29.20 & 40.99 & 6.81 & 25.79 & 0.0842 & 20.88 & 48.45 & 7.38 \\
\textbf{RIC (Interface)} & 15.55 & 23.80 & 33.72 & 5.78 & 21.14 & 0.0702 & 18.20 & 44.28 & 10.25 \\
\textbf{NAC (Antibody)} & \textbf{33.97} & \textbf{44.99} & \textbf{57.41} & \textbf{11.09} & \textbf{35.82} & \textbf{0.1791} & 34.49 & \textbf{69.14} & \textbf{1.38} \\
\tablemidrule
\multicolumn{10}{@{}l@{}}{\tablegroup{Protenix-v2}} \\
Pair-iPTM & 30.31 & 43.53 & 50.32 & 13.01 & 33.30 & 0.1674 & 33.77 & 63.50 & 6.88 \\
Pair-gPDE $\downarrow$ & 27.60 & 38.97 & 46.50 & 11.44 & 32.82 & 0.1529 & 27.72 & 60.23 & 11.38 \\
Interface PAE $\downarrow$ & 34.00 & 43.32 & 49.74 & \textbf{14.04} & 34.63 & 0.1866 & 33.80 & 64.35 & 4.25 \\
Antibody pLDDT & 32.94 & \underline{44.38} & \underline{52.92} & 12.95 & \underline{36.18} & 0.1837 & 34.44 & \underline{67.44} & \underline{3.50} \\
Ranking score & 29.82 & 44.15 & 51.60 & 13.64 & 33.52 & 0.1671 & 33.83 & 61.50 & 6.00 \\
pDockQ & 23.91 & 40.18 & 50.72 & 8.29 & 33.25 & 0.1280 & 30.49 & 57.61 & 10.62 \\
pDockQ2 & \underline{34.03} & 42.91 & 49.72 & \underline{13.74} & 34.15 & \underline{0.1914} & 34.37 & 64.03 & 4.62 \\
ipSAE-d0res & 30.15 & 42.93 & 50.58 & 13.08 & 33.37 & 0.1667 & 32.53 & 62.19 & 7.50 \\
ipSAE-d0dom & 30.96 & 43.21 & 49.69 & 13.15 & 33.30 & 0.1635 & 31.72 & 61.69 & 8.06 \\
ipSAE-d0chn & 31.95 & 41.87 & 49.75 & 13.15 & 32.59 & 0.1682 & 31.64 & 65.18 & 7.31 \\
AntiConf & 31.50 & 41.93 & 51.98 & 13.10 & 34.11 & 0.1792 & \underline{35.89} & 60.80 & 5.75 \\
\textbf{RIC (Interface)} & 14.28 & 23.31 & 33.42 & 5.46 & 18.65 & 0.0652 & 18.31 & 53.07 & 13.00 \\
\textbf{NAC (Antibody)} & \textbf{36.54} & \textbf{49.40} & \textbf{59.26} & 11.70 & \textbf{36.21} & \textbf{0.1983} & \textbf{37.77} & \textbf{67.59} & \textbf{2.12} \\
\tablemidrule
\multicolumn{10}{@{}l@{}}{\tablegroup{ESMFold2}} \\
Pair-iPTM & 24.60 & 33.95 & 46.83 & 8.92 & 33.61 & 0.1385 & 29.14 & 56.04 & 6.00 \\
Interface PAE $\downarrow$ & 25.50 & 34.09 & 46.59 & 9.23 & 33.80 & 0.1514 & 29.52 & 49.89 & 5.25 \\
Antibody pLDDT & \underline{32.78} & \underline{43.49} & \underline{51.40} & 10.19 & \underline{37.21} & \underline{0.1889} & 37.08 & \underline{58.40} & \underline{2.62} \\
pDockQ & 18.41 & 29.50 & 41.46 & 8.00 & 26.14 & 0.0934 & 22.94 & 47.79 & 9.25 \\
pDockQ2 & 24.79 & 34.50 & 44.92 & 9.38 & 33.78 & 0.1407 & 28.04 & 50.55 & 5.88 \\
ipSAE-d0res & 22.30 & 29.02 & 30.69 & 10.06 & 33.21 & 0.1210 & 39.56 & 52.86 & 7.38 \\
ipSAE-d0dom & 23.83 & 28.42 & 30.99 & \underline{10.44} & 33.89 & 0.1268 & \underline{39.81} & 54.32 & 5.75 \\
ipSAE-d0chn & 23.08 & 28.72 & 30.39 & 10.22 & 33.82 & 0.1299 & \textbf{40.11} & 57.02 & 5.75 \\
AntiConf & 24.63 & 33.62 & 44.02 & 9.63 & 33.38 & 0.1451 & 29.32 & 52.15 & 6.12 \\
\textbf{gRIC$^{*}$} & 11.88 & 22.53 & 35.15 & 4.69 & 21.19 & 0.0566 & 17.70 & 42.90 & 10.62 \\
\textbf{NAC (Antibody)} & \textbf{37.92} & \textbf{47.40} & \textbf{58.13} & \textbf{11.59} & \textbf{39.22} & \textbf{0.2165} & 39.45 & \textbf{62.96} & \textbf{1.38} \\
\tablemidrule
\multicolumn{10}{@{}l@{}}{\tablegroup{OpenDDE}} \\
Pair-iPTM & 23.60 & 34.96 & 43.44 & 9.55 & 30.97 & 0.1262 & 27.40 & 53.68 & 7.62 \\
Pair-gPDE $\downarrow$ & 23.45 & 33.06 & 43.40 & 9.23 & 30.87 & 0.1282 & 26.14 & 55.60 & 8.50 \\
Interface PAE $\downarrow$ & 22.17 & 33.99 & 44.66 & 9.10 & 30.88 & 0.1267 & 26.38 & 55.21 & 8.50 \\
Antibody pLDDT & \underline{25.42} & 36.21 & 45.30 & \textbf{11.24} & \underline{32.22} & \textbf{0.1404} & 28.79 & \textbf{59.25} & \underline{2.38} \\
Ranking score & 23.71 & 36.13 & \underline{46.56} & 10.33 & 31.34 & 0.1269 & 28.03 & 55.29 & 4.88 \\
pDockQ & 21.04 & 33.86 & 45.42 & 7.83 & 26.95 & 0.1021 & 27.82 & 49.95 & 9.75 \\
pDockQ2 & 23.42 & 33.84 & 42.74 & 9.88 & 31.48 & 0.1285 & 27.18 & 55.39 & 7.38 \\
ipSAE-d0res & 24.01 & 34.14 & 43.47 & 9.19 & 30.95 & 0.1318 & \underline{29.22} & 54.58 & 6.00 \\
ipSAE-d0dom & 22.90 & 34.13 & 44.59 & 8.71 & 31.30 & 0.1303 & \textbf{29.49} & 53.13 & 7.12 \\
ipSAE-d0chn & 20.97 & 31.00 & 43.63 & 8.78 & 29.09 & 0.1181 & 24.33 & 54.55 & 10.62 \\
AntiConf & 23.91 & \underline{36.87} & \textbf{46.63} & 10.42 & 32.17 & \underline{0.1353} & 27.30 & 56.65 & 3.25 \\
\textbf{RIC (Interface)} & 13.07 & 22.99 & 33.44 & 4.01 & 20.75 & 0.0647 & 20.84 & 47.34 & 13.00 \\
\textbf{NAC (Antibody)} & \textbf{25.82} & \textbf{37.17} & 45.54 & \underline{10.96} & \textbf{32.65} & 0.1336 & 29.06 & \underline{59.03} & \textbf{2.00} \\
\tablebottomrule
\end{xltabular}
\endgroup
\setcounter{table}{\value{resultstablebefore}}
\begin{figure}[H]
\centering
\begin{minipage}[t]{0.82\linewidth}
\vspace{0pt}
\centering
\begin{minipage}[t]{0.49\linewidth}
\vspace{0pt}
\includegraphics[width=\linewidth]{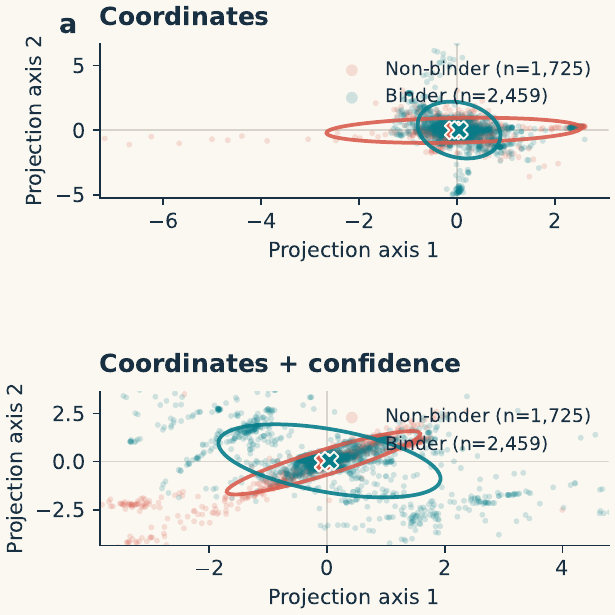}
\end{minipage}\hfill
\begin{minipage}[t]{0.49\linewidth}
\vspace{0pt}
\includegraphics[width=\linewidth]{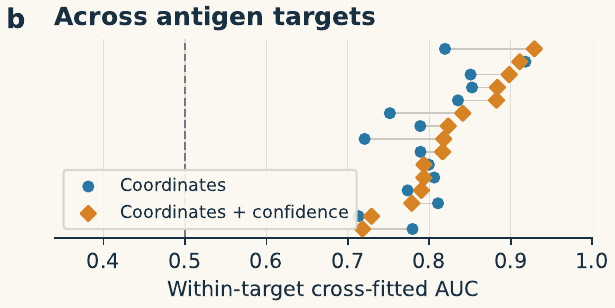}\par\vspace{0.4em}
\includegraphics[width=\linewidth]{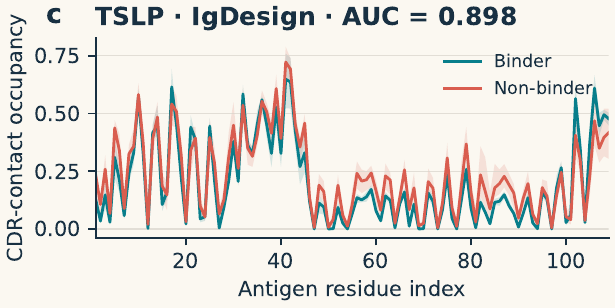}
\end{minipage}
\end{minipage}

\scriptsize
\caption{Binding information in AlphaFold~3 outputs.
\textbf{(a)} Projections of 4,184 candidates from coordinates (top) and
coordinates plus confidence (bottom), each over 10 seeds and 5 samples.
\textbf{(b)} Cross-fitted within-target AUCs across 15 antigen--dataset strata
with at least five binders and five non-binders; confidence improves
separability for many targets.
\textbf{(c)} Median CDR-contact occupancy (interquartile range) for 114 binders
and 49 non-binders against IgDesign TSLP. Contact regions largely overlap despite a
within-target AUC of 0.898.}
\label{fig:af3_output_distributions}
\label{fig:af3_epitope_localization}
\end{figure}

\subsection{Discussion}
\label{sec:discussion}

\paragraph{Predictor identifiability sets the computational ceiling.}
Here, identifiability refers to a simple limit: a downstream metric cannot
recover distinctions that are absent from the predictor outputs. In the
extreme, if binders and non-binders produce the same distribution of predicted
coordinates and confidence values, no deterministic score computed from those
outputs can distinguish them. Additional seeds and samples can reduce
stochastic variation, but they cannot correct systematic errors that are
reproduced across the ensemble. 

We first examine how much binding-related information is encoded in predicted
coordinates alone. Using AlphaFold~3 prediction ensembles, we represent each
candidate by coordinate-derived features, including CDR--antigen distances and
geometric contacts. A cross-fitted kernel witness analysis estimates binder and
non-binder separability within each antigen, achieving a target-macro AUC of
0.801. Thus, predicted coordinates contain substantial binding-related signal,
but the broad overlap between the two classes indicates that this information
is incomplete. The qualitative projection in
Figure~\ref{fig:af3_output_distributions}a illustrates this limited
separation.

Adding confidence tensors, including contact probabilities, bidirectional PAE,
and residue-level pLDDT, further improves separability (target-macro AUC from
0.801 to 0.827), showing that confidence tensors provide additional information
beyond coordinates. However, this improvement is not uniform across antigens:
some targets become substantially more separable, whereas others show little
gain or even reduced separation
(Figure~\ref{fig:af3_output_distributions}b).

For TSLP in the IgDesign benchmark, coordinate-derived contact patterns of
binders and non-binders overlap broadly across antigen regions
(Figure~\ref{fig:af3_output_distributions}c), suggesting that predicted
interaction locations alone are insufficient to explain binding outcomes.
Nevertheless, the complete structure-prediction outputs achieve a within-target
AUC of 0.898, indicating that additional information remains in confidence
tensors and interface-related properties. Together, these results highlight
both the opportunity and limit of structure-based filtering: better metrics can
extract more of the binding-related information encoded in predictor outputs,
but their ultimate performance is constrained by distinctions that the
predictors themselves do not capture.

\paragraph{Biological separability sets the benchmark ceiling.}
A benchmark is informative only when binder and non-binder labels represent
distinct biological outcomes for matched antibody--antigen constructs. Our
data combine display-based enrichment, replicated SPR, and literature-derived
affinity measurements, which differ in assay conditions, detection limits,
antigen formats, and binder definitions. Quality control reduces but cannot
eliminate this heterogeneity. The evaluation ceiling therefore also depends on
label fidelity.
Table~\ref{tab:separability_views} demonstrates that both display-based enrichment and
IgDesign show within-antigen AUCs significantly above chance for all three predictors.

\section{Conclusion and Limitations}

We revisit post-design antibody filtering as a practical selection problem under
limited experimental capacity, where candidates may be compared across
different antigens or ranked within a single antigen. We identify two
complementary signals from predicted complexes: interaction support for the
CDR--antigen interface, captured by RIC, and antibody structural confidence
outside the CDRs, captured by NAC. Across 4,184 labeled antibody--antigen pairs
and four structure predictors, we find that interface evidence is most useful
when candidates designed for different antigens compete, whereas antibody-level
confidence is more informative for ranking candidates within the same antigen.
Surprisingly, non-CDR confidence consistently outperforms whole-antibody pLDDT
for within-antigen ranking, revealing an underexplored source of filtering
signal beyond the predicted interface. More broadly, predicted geometry and
confidence provide useful but incomplete binding information: better metrics
can extract more from current predictors, but their ultimate performance is
limited by distinctions that the predictors themselves do not capture.

\noindent\textbf{Limitations.}
Only part of our benchmark comes from AI design; it also includes
display-selected antibodies and examples curated from the literature,
which may have different sequence and structural distributions
from AI-generated candidates. Still, these experimentally characterized antibodies allow us to
test whether co-folding outputs carry binding-related signals that can guide
filtering after AI design. Moreover, in the predictor training data available
for audit, we found no exact pre-cutoff antibody--antigen complexes, reducing
concern about direct memorization of the evaluated pairs.
The benchmark combines display-based enrichment,
replicated binding measurements, and published affinity data, which capture
related but not identical notions of binding.
Enrichment should be interpreted cautiously, because antibody
expression, clone growth, and sampling may also influence round-to-round
frequency changes.
Across all three predictors, the best cutoff varies by antigen
(Appendix~\ref{app:threshold_transfer}); predictor-derived scores should
therefore not be assumed to share a calibrated absolute scale across targets.
We optimize each metric's cutoff on labeled candidates for a fair benchmark
comparison. A new campaign instead needs independent calibration or a
prespecified top-$Q$ budget.
Limited computational resources restricted us to ten seeds and
five samples per seed; assessing whether more seeds and samples change our
conclusions remains for future work.

\newpage

{\small
\bibliographystyle{iclr2027_conference}
\bibliography{iclr2027_conference}
}

\newpage

\appendix

\section*{Appendix Contents}
\begingroup
\hypersetup{pdfborder={0 0 0}}
\setlength{\parindent}{0pt}
\newcommand{\appendixcontentsentry}[3]{%
  \par\addvspace{0.75em}\noindent
  \hyperref[#1]{\textbf{#2\quad #3}}\nobreak\dotfill\pageref*{#1}\par}
\newcommand{\appendixcontentssubentry}[3]{%
  \noindent\hspace*{2em}\hyperref[#1]{#2\quad #3}%
  \nobreak\dotfill\pageref*{#1}\par}
\appendixcontentsentry{app:additional_results}{A}{Additional Experimental Results}
\appendixcontentssubentry{app:results_by_benchmark}{A.1}{Matched population and benchmark results}
\appendixcontentssubentry{app:binder_separability}{A.2}{Binder separability and regional confidence}
\appendixcontentssubentry{app:low_prevalence}{A.3}{Filtering at lower binder prevalence}
\appendixcontentssubentry{app:ensemble_sensitivity}{A.4}{Sensitivity to ensemble size}
\appendixcontentssubentry{app:threshold_transfer}{A.5}{Threshold transfer to unseen antigens}
\appendixcontentsentry{app:benchmark_construction}{B}{Benchmark Construction}
\appendixcontentssubentry{app:yeast_display}{B.1}{Display-based enrichment benchmark}
\appendixcontentssubentry{app:igdesign_spr}{B.2}{IgDesign SPR benchmark}
\appendixcontentssubentry{app:literature_affinity}{B.3}{Literature and patent affinity benchmark}
\appendixcontentssubentry{app:assay_diversity}{B.4}{Assay definitions, constructs, and sequence diversity}
\appendixcontentsentry{app:experimental_details}{C}{Experimental Details}
\appendixcontentssubentry{app:cofolding_inference}{C.1}{Cofolding inference}
\appendixcontentssubentry{app:baseline_details}{C.2}{Baseline implementation}
\appendixcontentssubentry{app:campaign_specific_evaluation}{C.3}{Campaign-specific evaluation protocol}
\appendixcontentssubentry{app:threshold_selection}{C.4}{Threshold selection and fixed-budget evaluation}
\appendixcontentsentry{app:ric_implementation}{D}{RIC Implementation Details}
\appendixcontentsentry{sec:related}{E}{Related Work}
\endgroup
\hypersetup{pdfborder={0 0 1}}
\newpage

\section{Additional Experimental Results}
\label{app:additional_results}

\subsection{Matched population and benchmark results}
\label{app:results_by_benchmark}

Table~\ref{tab:matched_population_portfolio} repeats the global-ranking
evaluation on the same candidate population as Table~\ref{tab:target_specific_results}.
RIC/gRIC ranks ahead of NAC for all three predictors when candidates compete
across antigens. This reversal therefore persists without changing the
evaluated population. Each metric's global threshold is re-optimized on these
candidates, not carried over from Table~\ref{tab:global_results}. The matched
set excludes the mostly positive literature/patent cohort, which helps explain
its lower absolute results.

\begin{table}[!htbp]
\centering
\scriptsize
\setlength{\tabcolsep}{1.4pt}
\renewcommand{\arraystretch}{0.96}
\caption{\textbf{Global filtering on the candidates used in Table~\ref{tab:target_specific_results}.}
The 2,400 candidates (944 binders, 1,456 non-binders) from 13 antigens are
pooled and ranked with one shared threshold. This holds the candidate population
fixed while changing the comparison from within to across antigens. Average
ranks cover the seven reported measures. ESMFold2 excludes 107 out-of-memory
complexes; $^{*}$gRIC uses geometric contacts.}
\label{tab:matched_population_portfolio}
\begin{tabularx}{\textwidth}{@{}p{1.75cm}*{8}{>{\centering\arraybackslash}X}@{}}
\tabletoprule
\textbf{Metric} &
\shortstack{\textbf{Recall @}\\\textbf{10\% FPR}} &
\shortstack{\textbf{Recall @}\\\textbf{20\% FPR}} &
\shortstack{\textbf{Recall @}\\\textbf{30\% FPR}} &
\shortstack{\textbf{Recall @}\\$\boldsymbol{Q=5\%}$} &
\shortstack{\textbf{Recall @}\\$\boldsymbol{Q=20\%}$} &
\shortstack{\textbf{Utility}\\$\boldsymbol{U^\star}$} &
\shortstack{\textbf{Precision}\\\textbf{@ $U^\star$}} &
\shortstack{\textbf{Average}\\\textbf{rank}} \\
\tablemidrule
\multicolumn{9}{@{}l@{}}{\tablegroup{AlphaFold 3}} \\
Pair-iPTM & 2.97 & 7.20 & 17.06 & 2.05 & 12.29 & 0.0257 & 19.27 & 7.29 \\
Interface PAE $\downarrow$ & 3.50 & 6.14 & 12.08 & 2.97 & 9.85 & 0.0243 & 19.39 & 8.93 \\
Antibody pLDDT & 5.19 & 8.16 & 9.43 & 3.28 & 9.32 & 0.0355 & 23.93 & 4.93 \\
Ranking score & 2.75 & 6.14 & 11.12 & 3.43 & 10.14 & 0.0287 & 21.67 & 7.57 \\
pDockQ & 3.92 & 5.83 & 23.94 & 3.18 & 14.41 & 0.0312 & 28.33 & 4.29 \\
pDockQ2 & 3.81 & 6.25 & 12.18 & 2.86 & 10.38 & 0.0260 & 18.64 & 8.21 \\
ipSAE-d0res & 3.81 & 6.57 & 11.86 & 1.80 & 10.81 & 0.0257 & 19.27 & 8.21 \\
ipSAE-d0dom & 4.13 & 6.36 & 11.76 & 1.59 & 10.70 & 0.0272 & 20.79 & 7.57 \\
ipSAE-d0chn & 4.34 & 7.31 & 11.65 & 2.22 & 10.49 & 0.0280 & 21.99 & 6.29 \\
AntiConf & 3.81 & 6.14 & 13.45 & 2.86 & 11.12 & 0.0263 & 20.83 & 6.71 \\
\textbf{RIC (Interface)} & 6.04 & 7.52 & 13.45 & 3.28 & 10.81 & 0.0407 & 29.08 & \textbf{2.36} \\
\textbf{NAC (Antibody)} & 5.08 & 8.05 & 9.32 & 2.97 & 9.00 & 0.0382 & 24.70 & 5.64 \\
\tablemidrule
\multicolumn{9}{@{}l@{}}{\tablegroup{Protenix-v2}} \\
Pair-iPTM & 6.99 & 8.47 & 14.62 & 5.83 & 11.12 & 0.0498 & 40.00 & 2.93 \\
Pair-gPDE $\downarrow$ & 5.61 & 6.99 & 11.23 & 4.34 & 10.17 & 0.0380 & 24.51 & 10.64 \\
Interface PAE $\downarrow$ & 6.67 & 8.37 & 13.45 & 4.03 & 10.38 & 0.0464 & 34.07 & 7.07 \\
Antibody pLDDT & 6.99 & 8.37 & 10.81 & 4.45 & 9.53 & 0.0476 & 31.58 & 7.43 \\
Ranking score & 5.19 & 8.26 & 14.30 & 3.81 & 10.38 & 0.0346 & 26.63 & 9.93 \\
pDockQ & 6.25 & 7.31 & 10.17 & 4.24 & 8.47 & 0.0431 & 29.33 & 10.50 \\
pDockQ2 & 6.89 & 8.47 & 13.24 & 3.81 & 9.96 & 0.0473 & 32.02 & 7.64 \\
ipSAE-d0res & 5.30 & 6.78 & 14.09 & 4.45 & 10.38 & 0.0365 & 32.45 & 8.93 \\
ipSAE-d0dom & 6.78 & 8.47 & 14.19 & 5.61 & 10.59 & 0.0473 & 33.16 & 5.07 \\
ipSAE-d0chn & 6.99 & 8.47 & 15.04 & 6.25 & 11.23 & 0.0505 & 39.88 & 2.36 \\
AntiConf & 6.46 & 7.31 & 13.24 & 3.50 & 10.17 & 0.0436 & 29.25 & 9.93 \\
\textbf{RIC (Interface)} & 9.00 & 14.09 & 14.94 & 5.72 & 14.41 & 0.0710 & 36.30 & \textbf{1.86} \\
\textbf{NAC (Antibody)} & 6.14 & 8.58 & 16.10 & 3.28 & 10.81 & 0.0458 & 30.09 & 6.71 \\
\tablemidrule
\multicolumn{9}{@{}l@{}}{\tablegroup{ESMFold2}} \\
Pair-iPTM & 1.47 & 5.08 & 11.17 & 1.47 & 9.26 & 0.0250 & 18.26 & 10.86 \\
Interface PAE $\downarrow$ & 4.40 & 6.21 & 11.51 & 4.06 & 10.05 & 0.0290 & 46.75 & 6.36 \\
Antibody pLDDT & 5.87 & 8.80 & 9.71 & 4.29 & 9.59 & 0.0427 & 26.47 & 4.86 \\
pDockQ & 5.19 & 10.50 & 18.06 & 3.61 & 13.54 & 0.0406 & 24.91 & 3.43 \\
pDockQ2 & 4.18 & 5.87 & 11.40 & 4.06 & 10.05 & 0.0285 & 36.36 & 7.29 \\
ipSAE-d0res & 4.51 & 6.43 & 12.30 & 3.39 & 9.93 & 0.0292 & 22.40 & 7.50 \\
ipSAE-d0dom & 4.97 & 6.55 & 12.08 & 2.26 & 10.05 & 0.0324 & 25.00 & 6.43 \\
ipSAE-d0chn & 4.97 & 6.55 & 13.66 & 3.50 & 10.84 & 0.0328 & 25.73 & 5.14 \\
AntiConf & 4.18 & 5.98 & 15.58 & 4.06 & 9.93 & 0.0291 & 50.70 & 6.07 \\
\textbf{gRIC$^{*}$} & 7.00 & 10.05 & 12.98 & 4.18 & 11.29 & 0.0488 & 28.46 & \textbf{2.43} \\
\textbf{NAC (Antibody)} & 4.74 & 8.47 & 17.49 & 4.06 & 9.82 & 0.0329 & 22.12 & 5.64 \\
\tablebottomrule
\end{tabularx}
\end{table}

Table~\ref{tab:benchmark_campaign_results} similarly separates
the campaign-specific results in Table~\ref{tab:target_specific_results} by
benchmark. NAC leads on the display-based targets for all three predictors;
the IgDesign results are more mixed, particularly for Protenix-v2.

\begin{table}[!htbp]
\centering
\scriptsize
\setlength{\tabcolsep}{2.5pt}
\renewcommand{\arraystretch}{1.08}
\caption{\textbf{Campaign-specific results from
Table~\ref{tab:target_specific_results}, separated by benchmark.} Antigens are
evaluated separately, then aggregated within each benchmark. R10 is recall at
10\% FPR (in percent), $U$ is filtering utility, and ranks use all seven
measures. IgDesign includes six eligible antigens; ESMFold2 excludes 107
out-of-memory display-based candidates.}
\label{tab:benchmark_campaign_results}
\begin{tabularx}{\textwidth}{@{}p{1.7cm}p{2.0cm}p{1.5cm}XXp{1.35cm}@{}}
\tabletoprule
\textbf{Predictor} & \textbf{Candidates (P/N)} & \textbf{Best metric} &
\shortstack{\textbf{RIC/gRIC}\\\textbf{rank / R10 / $U$}} &
\shortstack{\textbf{NAC}\\\textbf{rank / R10 / $U$}} &
\shortstack{\textbf{Antibody}\\\textbf{pLDDT rank}} \\
\tablemidrule
\multicolumn{6}{@{}l@{}}{\tablegroup{Display-based enrichment}} \\
AlphaFold 3 & 1,400 (700/700) & NAC & 5.64 / 18.43 / 0.0898 & 1.07 / 30.00 / 0.1626 & 3.07 \\
Protenix-v2 & 1,400 (700/700) & NAC & 6.71 / 15.29 / 0.0722 & 1.07 / 29.00 / 0.1430 & 2.21 \\
ESMFold2 & 1,293 (642/651) & NAC & 4.71 / 8.48 / 0.0378 & 1.00 / 28.53 / 0.1433 & 2.14 \\
\tablemidrule
\multicolumn{6}{@{}l@{}}{\tablegroup{IgDesign SPR}} \\
AlphaFold 3 & 1,000 (244/756) & NAC & 11.57 / 12.68 / 0.0506 & 1.00 / 37.94 / 0.1957 & 2.43 \\
Protenix-v2 & 1,000 (244/756) & Interface PAE & 13.00 / 13.28 / 0.0583 & 9.57 / 44.09 / 0.2537 & 7.57 \\
ESMFold2 & 1,000 (244/756) & NAC & 11.00 / 15.29 / 0.0753 & 1.57 / 47.32 / 0.2897 & 4.14 \\
\tablebottomrule
\end{tabularx}
\end{table}

\Needspace{0.55\textheight}
\subsection{Binder separability and regional confidence}
\label{app:binder_separability}

Table~\ref{tab:separability_views} examines how well the
predicted outputs distinguish binders from non-binders within an antigen.
Confidence yields higher mean AUC than coordinates alone in all six
benchmark--predictor comparisons, although combining the two views does not
always improve it further.

\begin{table}[H]
\centering
\small
\setlength{\tabcolsep}{8pt}
\renewcommand{\arraystretch}{1.08}
\caption{\textbf{Within-antigen binder separability from predicted coordinates and confidence.}
Values are unweighted mean AUCs over the seven antigens in each benchmark,
using five-fold within-antigen cross-fitting. All three predictors are evaluated
on the same complete-ensemble candidates (1,293 display-based and 1,171
IgDesign pairs). The combined view joins coordinates and confidence. These
views include non-CDR antibody pLDDT and omit predicted contact probabilities,
so they differ from those in Figure~\ref{fig:af3_output_distributions}. The matched
literature/patent subset has no antigen with enough examples of both classes
for within-antigen AUC.}
\label{tab:separability_views}
\begin{tabular}{@{}llccc@{}}
\tabletoprule
\textbf{Benchmark} & \textbf{Predictor} & \textbf{Coordinates} & \textbf{Confidence} & \textbf{Combined} \\
\tablemidrule
Display-based enrichment & AlphaFold 3 & 0.811 & 0.855 & 0.847 \\
 & Protenix-v2 & 0.808 & 0.810 & 0.808 \\
 & ESMFold2 & 0.779 & 0.822 & 0.820 \\
\tablemidrule
IgDesign SPR & AlphaFold 3 & 0.796 & 0.822 & 0.823 \\
 & Protenix-v2 & 0.817 & 0.828 & 0.831 \\
 & ESMFold2 & 0.825 & 0.829 & 0.838 \\
\tablebottomrule
\end{tabular}
\end{table}

\begingroup
\paragraph{Regional antibody confidence.}
The CDRs and binding interface are central to recognition, but their
conformations and contacts can also be difficult to predict
\citep{unsal2026anticonf,aintibody2026benchmark}. Whole-antibody pLDDT mixes
framework and CDR confidence, while interface pLDDT also depends on
predicted docking. This suggests a trade-off between direct relevance to
binding and prediction reliability. Figure~\ref{fig:regional_confidence_comparison}
examines seven regional summaries, linking binder separability to
fixed-budget recall and threshold-based utility.

\paragraph{Which residues contribute?}
NAC excludes all CDRs, whereas $\mathrm{N}^{3}\mathrm{AC}$ excludes only
CDR3, retaining CDR1, CDR2, and non-CDR residues. Conversely, CDR3+NAC
retains CDR3 and non-CDR residues, excluding CDR1/CDR2. CDR and
CDR3-only average over all CDRs and CDR3 alone, respectively. These
definitions use IMGT assignments on both heavy and light chains when
present. Residues outside the annotated variable domains remain in the
non-CDR summaries. Interface pLDDT averages unique residues on the antibody
variable domains and antigen that form predicted 8-\AA{} C$\beta$ contacts
(C$\alpha$ when C$\beta$ is unavailable), while samples without contacts score zero.
Every score averages residue pLDDT within a sample, then takes the median
over five samples per seed and over ten seeds. Only the confidence summary
changes: all predictions co-fold the full antibody, including every CDR,
with its antigen.

For each antigen, we compute AUC directly from the scores, recall at a
fixed $Q=20\%$ budget, and maximum utility using the same threshold sweep
as in the main evaluation. Each benchmark then gives equal weight to its
seven antigens. All four predictors use the same 2,464 candidates
(1,293 display-based and 1,171 IgDesign pairs). 

\begin{figure}[p]
\centering
\includegraphics[width=\linewidth]{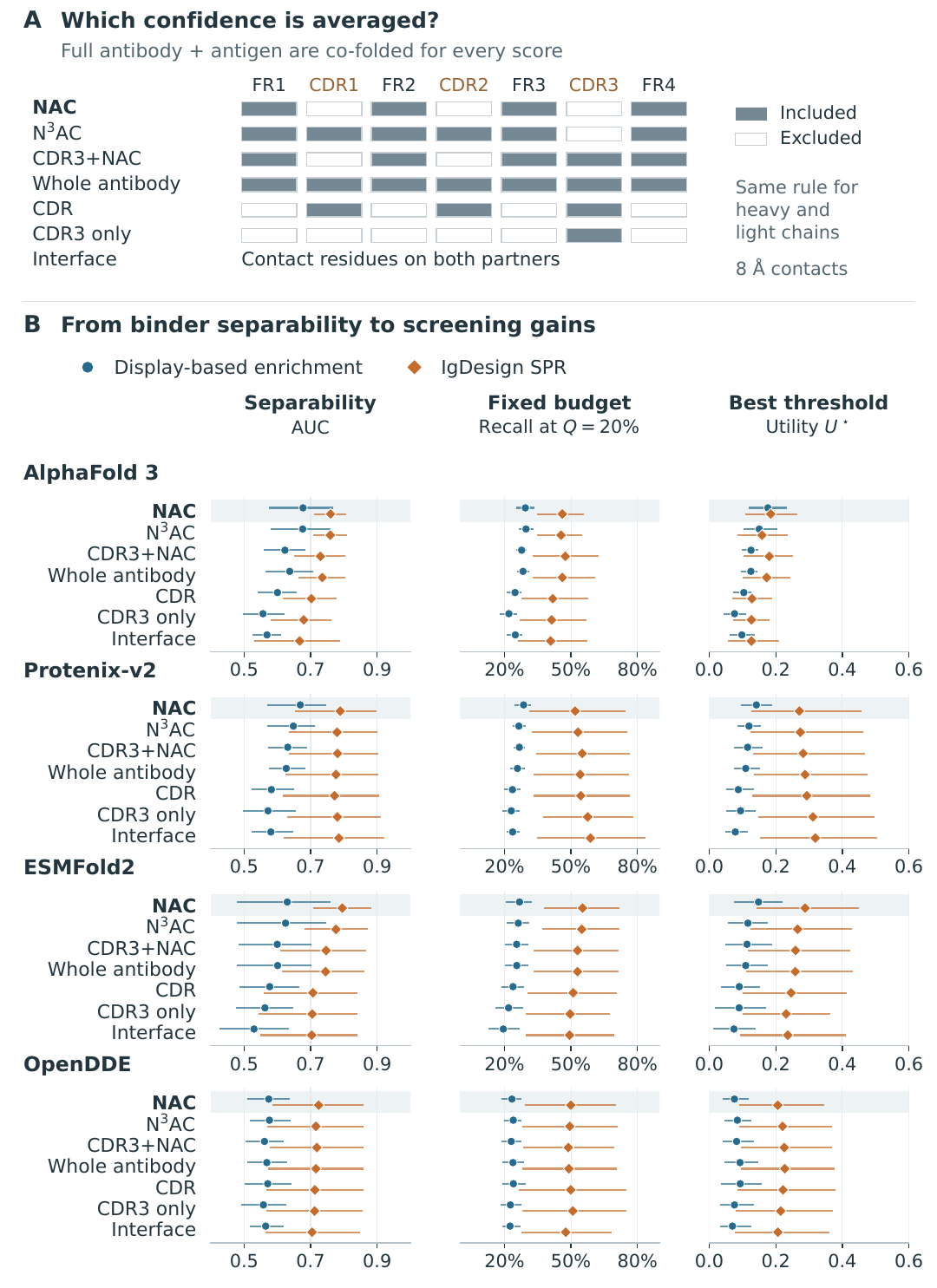}
\caption{\textbf{Regional confidence: from binder separability
to screening gains.} (A) Included antibody regions; the interface additionally
includes antigen contact residues. (B) Equal-antigen means for seven antigens
per benchmark. Blue circles denote display-based enrichment; orange diamonds
denote IgDesign SPR. Lines show 95\% intervals from 10,000 paired antigen
bootstrap resamples, shared across scores and predictors. Dashed lines mark
AUC $=0.5$ and nominal random-selection recall of $20\%$. Columns share their
axis ranges across predictors; shading identifies NAC. Utility selects the
best threshold using labels, whereas recall uses a prespecified budget.
Intervals reflect antigen variation, not independent validation of the
selected thresholds.}
\label{fig:regional_confidence_comparison}
\end{figure}

\paragraph{Separability and screening gains.}
NAC has the highest mean AUC in seven of eight predictor--benchmark comparisons.
Excluding CDR3 alone improves AUC over whole-antibody confidence in all eight.
For OpenDDE on display-based enrichment, $\mathrm{N}^{3}\mathrm{AC}$ slightly
exceeds NAC (0.576 versus 0.574). Comparing NAC, whole-antibody, CDR, and
interface confidence, AUC follows that order in six of eight comparisons;
the exceptions are OpenDDE on display-based enrichment and Protenix-v2 on
IgDesign. One possible explanation for the latter is that Protenix-v2 models
the interfaces of these IgDesign candidates more reliably, making interface
confidence particularly useful for prioritizing binders within these antigens.

NAC has the highest utility in five of eight comparisons, but the highest fixed-budget
recall in three. For AlphaFold~3, $\mathrm{N}^{3}\mathrm{AC}$ slightly leads
recall on display-based enrichment (29.7\% versus 29.4\% for NAC), and
CDR3+NAC leads on IgDesign (47.5\% versus 46.1\%). Protenix-v2 on IgDesign
shows a clearer reversal: interface confidence gives the highest recall
(58.8\%) and utility (0.319), despite NAC having the highest AUC
(0.789 versus 0.785). CDR3-only confidence also performs well in this setting,
but is weaker than NAC on all three endpoints in the other five comparisons
among AlphaFold~3, Protenix-v2, and ESMFold2. OpenDDE adds further exceptions:
on display-based enrichment, CDR confidence leads recall (24.1\%) and utility
(0.093); on IgDesign, CDR3-only leads recall (50.9\%), while whole-antibody
confidence leads utility (0.227). Thus, the best region for AUC need not be
the best for a particular filtering objective.

\textbf{Our observation.} Excluding CDR confidence strengthens within-antigen
binder separability across predictors and often improves screening gains,
but the most useful region also depends on the predictor, benchmark, and
selection objective. This pattern is consistent with the proposed trade-off:
a region's relevance to binding need not determine the usefulness of its
predicted confidence as a filtering signal.

\endgroup

\subsection{Filtering at lower binder prevalence}
\label{app:low_prevalence}
New AI-design candidate pools may contain a smaller fraction of binders than
the benchmark. To examine this setting, we resampled candidates within each
antigen to approximately 5\%, 10\%, 15\%, and 20\% binder prevalence, then
repeated fixed-budget filtering. The analysis uses the 1,935 candidates from
six display-based and four IgDesign antigens with complete predictions from
all three predictors and enough candidates to retain at least five binders
per antigen at 5\% prevalence. We repeated each resampling 500 times and used
the same sampled candidates for every score. Table~\ref{tab:low_prevalence}
compares NAC with the strong antibody pLDDT baseline from Table~\ref{tab:target_specific_results}.

\begin{table}[H]
\centering
\small
\setlength{\tabcolsep}{5pt}
\renewcommand{\arraystretch}{1.10}
\caption{\textbf{Within-antigen filtering at lower binder prevalence.}
Each predictor column reports mean Precision / Recall (\%) at $Q=20\%$ over
500 resamples; random selection is evaluated with the same per-antigen budget.
Binders or non-binders are subsampled within each antigen to reach the stated
prevalence. The 10 antigen weights are fixed from the matched population
before subsampling, with equal total weight for the two benchmarks.}
\label{tab:low_prevalence}
\begin{tabular}{@{}clcccc@{}}
\tabletoprule
\textbf{Binder rate} & \textbf{Score} & \textbf{Random} & \textbf{AlphaFold 3} & \textbf{Protenix-v2} & \textbf{ESMFold2} \\
\tablemidrule
\multirow{2}{*}{5\%} & NAC & \multirow{2}{*}{4.9 / 19.8} & 11.3 / 45.6 & 12.7 / 51.2 & 13.6 / 54.2 \\
 & Antibody pLDDT & & 11.2 / 44.8 & 11.6 / 46.6 & 12.8 / 50.8 \\
\tablemidrule
\multirow{2}{*}{10\%} & NAC & \multirow{2}{*}{10.0 / 19.8} & 22.6 / 44.9 & 24.6 / 48.9 & 26.9 / 53.4 \\
 & Antibody pLDDT & & 21.4 / 42.6 & 23.3 / 46.3 & 25.6 / 50.8 \\
\tablemidrule
\multirow{2}{*}{15\%} & NAC & \multirow{2}{*}{15.1 / 19.6} & 33.3 / 43.5 & 35.4 / 46.3 & 38.8 / 50.9 \\
 & Antibody pLDDT & & 30.9 / 40.5 & 34.6 / 45.3 & 37.3 / 48.9 \\
\tablemidrule
\multirow{2}{*}{20\%} & NAC & \multirow{2}{*}{20.0 / 20.0} & 42.7 / 42.7 & 44.3 / 44.3 & 49.1 / 49.1 \\
 & Antibody pLDDT & & 40.0 / 40.0 & 42.6 / 42.6 & 48.0 / 48.0 \\
\tablebottomrule
\end{tabular}
\end{table}

At 5\% binder prevalence, NAC recovers 45.6--54.2\% of binders while
testing 20\% of candidates, compared with 19.8\% recall under random selection.
Its precision is 11.3--13.6\%, more than twice the random expectation of
4.9\%. NAC also exceeds whole-antibody pLDDT in mean recall at every
predictor--prevalence combination shown. These subsets change class balance
among observed candidates; they do not reproduce the sequence distribution
of a new AI-design campaign.

\Needspace{0.82\textheight}
\subsection{Sensitivity to ensemble size}
\label{app:ensemble_sensitivity}
We tested whether smaller prediction ensembles preserve
fixed-budget filtering performance. On the 2,400 campaign candidates with
complete AlphaFold 3 and Protenix-v2 predictions, we compared two, six, and
ten runs (seeds), retaining five samples per run. We also fixed ten runs and
compared one, three, and five samples per run. RIC, NAC, antibody pLDDT,
pair-iPTM, and ipSAE-d0chn were recomputed from the same prediction subsets
before evaluating the same candidates and labels. The full ten-run,
five-sample scores reproduce Table~\ref{tab:target_specific_results}.

\begin{figure}[H]
\centering
\includegraphics[width=0.91\textwidth]{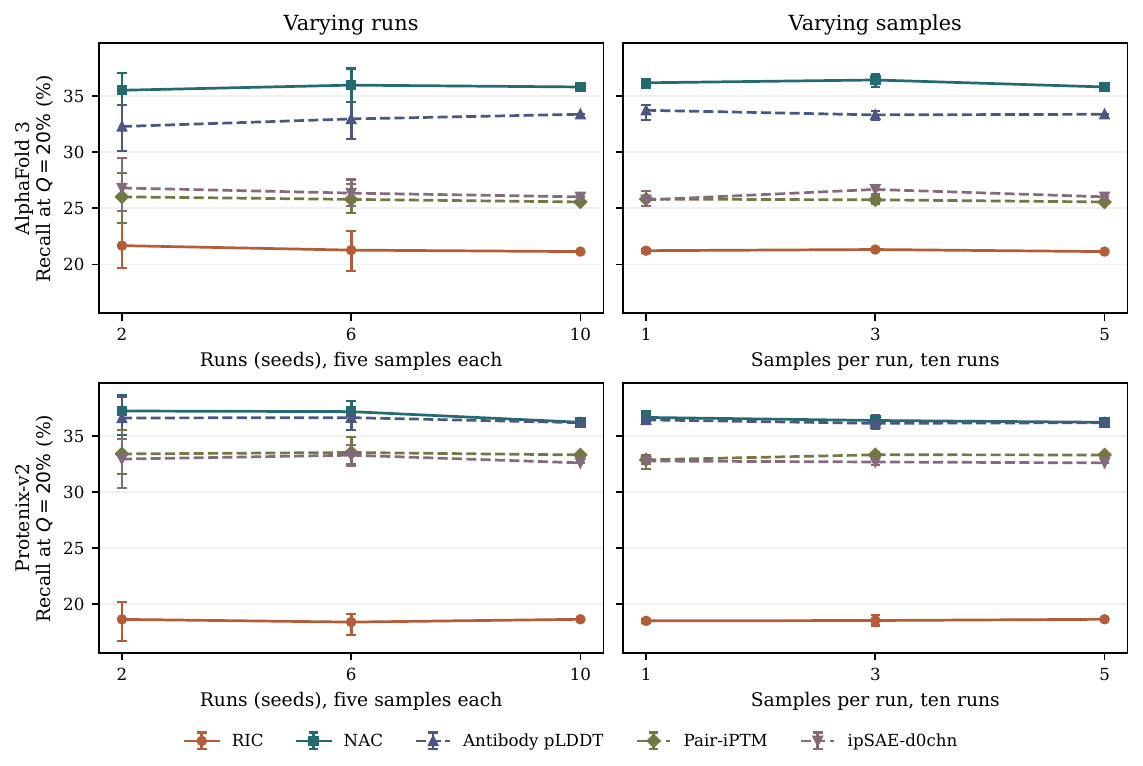}
\caption{\textbf{Fixed-budget recall with smaller prediction
ensembles.} Within-antigen Recall at $Q=20\%$ for AlphaFold 3 (top) and
Protenix-v2 (bottom). Points are means and bars span the selected subsets.
The run analysis uses all 45 two-run pairs and 50 prespecified six-run
subsets; the sample analysis uses three prespecified subsets at each reduced
count. The full ensemble has ten runs and five samples per run. All curves
use the same 2,400 campaign candidates; ESMFold2 is not included.}
\label{fig:ensemble_sensitivity}
\end{figure}

Figure~\ref{fig:ensemble_sensitivity} shows that NAC retains
the highest mean Recall at $Q=20\%$ among these scores at every tested
budget. With two versus ten runs, its mean recall is 35.5\% versus 35.8\%
for AlphaFold 3 and 37.2\% versus 36.2\% for Protenix-v2. Mean recall also
changes little when the number of samples per run varies. Individual run
choices matter, however: with two runs, RIC recall ranges from 19.7\% to
23.7\% for AlphaFold 3 and from 16.7\% to 20.2\% for Protenix-v2. Thus,
smaller ensembles preserve much of the average fixed-budget signal on these
candidates, but their rankings and threshold-dependent performance need not
be identical.

\Needspace{0.78\textheight}
\subsection{Threshold transfer to unseen antigens}
\label{app:threshold_transfer}
Table~\ref{tab:threshold_transfer} tests whether a threshold
chosen on labeled antigens can be applied to a new antigen. We hold out each
of the 13 campaign antigens in turn. For each predictor and score, we choose
one absolute cutoff that maximizes the weighted mean utility on the other
12 antigens, then apply it unchanged to the held-out antigen. We also calibrate
using only the other antigens from the same assay benchmark. The held-out
labels are used only to evaluate the selected cutoff. Both calibration
strategies use the score directions, threshold tie rules, and utility
definition of Table~\ref{tab:target_specific_results}.

\begin{table}[H]
\centering
\begingroup
\small
\setlength{\tabcolsep}{6pt}
\renewcommand{\arraystretch}{1.04}
\caption{\textbf{Leave-one-antigen-out threshold transfer.} Weighted mean
held-out utility when the cutoff is selected on all other antigens or only
on other antigens from the same assay benchmark. The oracle column uses
each held-out antigen's own labels to select its best cutoff and repeats
$U^\star$ from Table~\ref{tab:target_specific_results}; it is not a
prospective estimate. The final column counts held-out antigens on which
the all-other-antigen cutoff selects at least one candidate, out of 13.
AlphaFold 3 and Protenix-v2 use 2,400 candidates each; ESMFold2 uses its
2,293 successfully predicted candidates.}
\label{tab:threshold_transfer}
\begin{tabular}{@{}llrrrr@{}}
\tabletoprule
Predictor & Score & All other & Same assay & Oracle & Selected / 13 \\
\tablemidrule
\multirow{5}{*}{AlphaFold 3}
 & RIC & 0.0490 & 0.0503 & 0.0702 & 12 \\
 & NAC & 0.0000 & 0.0015 & 0.1791 & 1 \\
 & Antibody pLDDT & 0.0138 & 0.0246 & 0.1367 & 11 \\
 & Pair-iPTM & 0.0548 & 0.0427 & 0.0807 & 12 \\
 & ipSAE-d0chn & 0.0122 & 0.0137 & 0.0539 & 10 \\
\tablemidrule
\multirow{5}{*}{Protenix-v2}
 & RIC & 0.0322 & 0.0300 & 0.0652 & 13 \\
 & NAC & 0.0085 & 0.0423 & 0.1983 & 2 \\
 & Antibody pLDDT & 0.0185 & 0.0412 & 0.1837 & 2 \\
 & Pair-iPTM & 0.0920 & 0.1197 & 0.1674 & 4 \\
 & ipSAE-d0chn & 0.0837 & 0.0993 & 0.1682 & 4 \\
\tablemidrule
\multirow{5}{*}{ESMFold2}
 & gRIC & 0.0080 & 0.0069 & 0.0566 & 9 \\
 & NAC & 0.0047 & 0.0086 & 0.2165 & 2 \\
 & Antibody pLDDT & 0.0133 & 0.0475 & 0.1889 & 2 \\
 & Pair-iPTM & 0.0000 & 0.0029 & 0.1385 & 2 \\
 & ipSAE-d0chn & 0.0012 & 0.0043 & 0.1299 & 5 \\
\tablebottomrule
\end{tabular}
\endgroup
\end{table}

The strong within-antigen NAC results do not imply that one
absolute NAC cutoff transfers across antigens. Calibration on all other
antigens selects candidates on only one or two held-out antigens, depending
on the predictor, and its held-out utility is well below the per-antigen
oracle. Reusing an absolute cutoff tests whether score scales are comparable
across antigens, much like across-antigen filtering. RIC/gRIC retains a
larger share of its per-antigen oracle utility than NAC for all three
predictors, consistent with its relative strength across antigens. This
does not diminish NAC's within-antigen ranking result:
\textbf{within-antigen candidate ranking can remain informative even when
absolute NAC scores shift between targets}.
Limiting calibration to the same assay benchmark improves several scores,
including Protenix-v2 NAC, but leaves a substantial gap. \textbf{Our observation:}
The best absolute threshold depends on the antigen for every metric tested;
cutoffs calibrated on other antigens fall short of per-antigen optimization.
This may reflect
antigen-dependent score biases, potentially shaped by the predictors' training
data. Fixed-budget top-$Q$ selection avoids an absolute cutoff while retaining
NAC's strong within-antigen ranking signal: at $Q=20\%$, it recovers
35.82\%, 36.21\%, and 39.22\% of binders with AlphaFold~3, Protenix-v2, and
ESMFold2, respectively (Table~\ref{tab:target_specific_results}).

\section{Benchmark Construction}
\label{app:benchmark_construction}

\subsection{Display-based enrichment benchmark}
\label{app:yeast_display}

\paragraph{Experimental readout.}
The benchmark combines two display platforms. Phage display
was used for CD98hc, EGFR, PLXNA1 ectodomain, and TNFRSF23. Antibody fragments
displayed on phage were incubated with antigen-coated magnetic beads; retained
phage were recovered, amplified, and carried into the next selection round.
Yeast display was used for TfR, SARS-CoV-2 spike S2, and GPC3. Antibody
fragments displayed on yeast were selected with labeled antigen using magnetic
and fluorescence-activated sorting~\citep{chao2006yeastdisplay,huang2022yeastngs}.
Both platforms were sequenced across rounds to track clone enrichment.

For the four phage-display targets, 5, 4, 2, and 1~$\mu$g of antigen were used
in successive panning rounds. For the three yeast-display targets, the final
sorting round used antigen at 1~nM.

\paragraph{Filtering, quality control, and label construction.}
For each antigen, reads from all available rounds are translated, sequence
quality filtered, exact-deduplicated, and grouped by antibody sequence. For
clone $a$ in round $r$, its normalized abundance is
$f_{a,r}=n_{a,r}/\sum_b n_{b,r}$, where $n_{a,r}$ is the number of retained
reads for clone $a$. We calculate an enrichment signal from the normalized
frequency change between the early and later selection stages,
$E_a=\log_2[(f_{a,\mathrm{late}}+\varepsilon)/(f_{a,\mathrm{early}}+\varepsilon)]$,
where $\varepsilon$ is a fixed read-level pseudocount. Candidates must pass
sequence and read-quality filters and have a frequency trajectory consistent
with the sign of $E_a$. For each antigen, the up to 100 late-enriched candidates
with the largest $E_a$ form the positive set. Negative controls are drawn from
the baseline or earliest-round pool and show depletion or no enrichment in later
rounds; early low-enrichment candidates are used only when needed. Later rounds
use lower antigen inputs, so $E_a$ reflects survival under increasing
selection stringency.

\paragraph{Composition.}
The display-based enrichment benchmark contains 1,400 antibodies across seven antigens,
with 100 positives and 100 negatives for each antigen. The included antigen
systems are CD98hc, EGFR, GPC3, PLXNA1 ectodomain, SARS-CoV-2 spike S2,
transferrin receptor, and TNFRSF23.

\paragraph{Library depth and sequence diversity.}
Table~\ref{tab:yeast_library_depth} reports the scale of the source libraries
before benchmark subsampling. For the six experiments supplied as per-sequence
count tables, an input record is one sequence entry carried into AIRR/full-VH
processing.
``QC-retained read count'' is the sum of duplicate-aware read counts after
productive, in-frame IGH and amino-acid quality control. ``Unique full VH'' is
the number of exact full-length VH amino-acid sequences retained within that
round. Consequently, the antigen-level total in the final column counts
sequence--round observations rather than deduplicating a clone observed in
multiple rounds. Across the seven antigens, individual rounds contain
56,284--1,878,482 unique full-VH sequences and approximately
$2.34\times10^5$--$4.57\times10^6$ retained reads, substantially exceeding the
200 candidates per antigen included in the benchmark. In aggregate, the source
libraries comprise 26,540,032 input sequence records, a QC-retained read count
of 36,235,220, and 15,226,144 unique sequence--round observations.

\begin{longtable}{@{}llcrrr@{}}
\caption{\textbf{Source-library depth for the display-based enrichment benchmark.} Input sequence
records are records entering full-VH quality control. QC-retained read count sums
duplicate-aware counts after quality control; unique full VH denotes exact
amino-acid sequences within a round. Antigen-level totals sum round-level
values, so the final-column total is a count of sequence--round observations,
not an across-round deduplicated clone count. A dash marks the pre-selection
baseline, a total, or a yeast-display round for which the antigen concentration
was not provided. Phage-display inputs are antigen masses; yeast-display inputs
are antigen concentrations.}
\label{tab:yeast_library_depth}\\
\tabletoprule
Antigen & Round & \makecell{Antigen\\input} &
\makecell[r]{Input sequence\\records} &
\makecell[r]{QC-retained\\read count} &
\makecell[r]{Unique\\full VH} \\
\tablemidrule
\endfirsthead
\multicolumn{6}{l}{\textit{Table~\ref{tab:yeast_library_depth} continued}}\\
\tabletoprule
Antigen & Round & \makecell{Antigen\\input} &
\makecell[r]{Input sequence\\records} &
\makecell[r]{QC-retained\\read count} &
\makecell[r]{Unique\\full VH} \\
\tablemidrule
\endhead
\tablemidrule
\multicolumn{6}{r}{\textit{Continued on next page}}\\
\endfoot
\tablebottomrule
\endlastfoot
CD98hc & R1 & 5~$\mu$g & 3,019,877 & 2,237,462 & 1,878,482 \\
       & R2 & 4~$\mu$g &   288,646 &   247,728 &   162,207 \\
       & R3 & 2~$\mu$g & 2,085,258 & 2,095,645 & 1,666,243 \\
       & R4 & 1~$\mu$g &   417,666 & 1,845,768 &   126,526 \\
       & Total & -- & 5,811,447 & 6,426,603 & 3,833,458 \\
\tablemidrule
EGFR   & R1 & 5~$\mu$g & 1,169,527 &   924,008 &   689,821 \\
       & R2 & 4~$\mu$g & 1,243,577 & 1,576,915 &   774,560 \\
       & R3 & 2~$\mu$g & 2,503,238 & 4,570,804 & 1,296,785 \\
       & R4 & 1~$\mu$g & 1,276,218 & 2,472,962 &   605,591 \\
       & Total & -- & 6,192,560 & 9,544,689 & 3,366,757 \\
\tablemidrule
GPC3   & R1 & -- &   705,045 &   542,241 &   324,749 \\
       & R2 & -- &   800,666 &   903,710 &   436,174 \\
       & R3 & -- &   166,630 &   389,192 &    56,284 \\
       & R4 & 1~nM &   951,617 &   632,421 &   533,960 \\
       & Total & -- & 2,623,958 & 2,467,564 & 1,351,167 \\
\tablemidrule
PLXNA1 ECD & R1 & 5~$\mu$g & 1,315,443 &   802,537 &   676,591 \\
           & R2 & 4~$\mu$g &   954,430 & 1,092,403 &   613,320 \\
           & R3 & 2~$\mu$g &   863,064 & 1,153,048 &   500,881 \\
           & R4 & 1~$\mu$g &   813,671 & 1,247,029 &   414,013 \\
           & Total & -- & 3,946,608 & 4,295,017 & 2,204,805 \\
\tablemidrule
Spike S2 & R1 & -- &   950,482 & 1,816,251 &   499,915 \\
         & R2 & -- &   255,387 &   905,852 &   109,195 \\
         & R3 & -- &   562,782 &   233,744 &   132,920 \\
         & R4 & 1~nM &   366,993 & 1,541,237 &   134,778 \\
         & Total & -- & 2,135,644 & 4,497,084 &   876,808 \\
\tablemidrule
TfR    & R0 & -- &   127,668 &   445,113 &   100,009 \\
       & R1 & -- &   651,716 &   926,675 &   500,796 \\
       & R2 & -- &   550,195 & 1,272,308 &   343,741 \\
       & R3 & 1~nM &   577,530 & 1,287,790 &   346,152 \\
       & Total & -- & 1,907,109 & 3,931,886 & 1,290,698 \\
\tablemidrule
TNFRSF23 & R1 & 5~$\mu$g &   966,612 &   742,951 &   512,726 \\
         & R2 & 4~$\mu$g & 1,344,649 & 1,709,197 &   912,832 \\
         & R3 & 2~$\mu$g &   925,774 & 1,470,200 &   534,134 \\
         & R4 & 1~$\mu$g &   685,671 & 1,150,029 &   342,759 \\
         & Total & -- & 3,922,706 & 5,072,377 & 2,302,451 \\
\tablemidrule
All antigens & Total & -- & 26,540,032 & 36,235,220 & 15,226,144 \\
\end{longtable}

\subsection{IgDesign SPR benchmark}
\label{app:igdesign_spr}

\paragraph{Source experiment.}
IgDesign conditions antibody inverse folding on a native antibody--antigen
complex backbone and framework sequence, then redesigns either HCDR3 or all
three heavy-chain CDRs. The original study generated 100 HCDR3 and 100 HCDR123
designs for each of eight therapeutic antigens and screened them by
SPR~\citep{shanehsazzadeh2023igdesign}. Seven unblinded antigen systems have
sufficient released sequence and replicate-level binding data for this study.

\paragraph{Replicate-consistent labels and sequence reconstruction.}
We retain a designed antibody as positive only when every available replicate
reports binding and as negative only when no replicate reports binding; both
classes require at least two replicates. Positive-control antibodies and rows
with discordant or incomplete replicate calls are excluded. The released
$K_D$ value is the mean of quantifiable replicate measurements; it is not the
threshold used for the binary binder label. Full variable
domains are reconstructed by grafting the released designed HCDRs into the
corresponding reference scaffold, while preserving the matched light chain and
antigen construct. The resulting benchmark contains 249 binders and 922
non-binders.

\subsection{Literature and patent affinity benchmark}
\label{app:literature_affinity}

\paragraph{Source collection and affinity labels.}
We extract antibody sequences, antigen identities, experimental constructs,
and direct equilibrium affinity measurements from primary papers and patents.
Positive labels require a monovalent SPR, biolayer interferometry (BLI), or
isothermal titration calorimetry (ITC) measurement with $K_D\leq10$~nM;
negative labels require $K_D\geq1{,}000$~nM. EC$_{50}$- or IC$_{50}$-only
measurements, cell-based readouts, competition-only experiments, and ambiguous
or conflicting records are excluded.

\paragraph{Construct matching.}
Each retained record must support the antibody variable-domain sequence and an
antigen construct suitable for cofolding. We prefer the exact experimental
antigen sequence. When the source specifies a standard extracellular domain,
receptor-binding domain, soluble construct, or explicit residue range without
printing its full sequence, we map that construct to the corresponding
reference sequence. Records lacking a defensible construct assignment are
discarded. The final benchmark contains 1,613 pairs: 1,510 binders and 103
non-binders across 165 antigen identities. The
literature/patent benchmark contributes to portfolio selection but is excluded
from campaign selection because its antigen-level negative support is too
sparse for the reported low-FPR endpoints.

\paragraph{Per-antigen composition.}
Table~\ref{tab:target_composition} reports the 13 antigens in
the campaign-selection population. These comprise all seven display-based enrichment
antigens and the six IgDesign antigens that satisfy the minimum positive and
negative support requirements.

\begin{longtable}{@{}>{\raggedright\arraybackslash}p{3.2cm}>{\raggedright\arraybackslash}p{5.3cm}rr@{}}
\caption{Target-level composition of the display-based enrichment and
IgDesign benchmarks used for campaign selection.}
\label{tab:target_composition}\\
\tabletoprule
Benchmark & Antigen & Positive & Negative \\
\tablemidrule
\endfirsthead
\multicolumn{4}{c}{\tablename\ \thetable\ continued}\\
\tabletoprule
Benchmark & Antigen & Positive & Negative \\
\tablemidrule
\endhead
\tablemidrule
\multicolumn{4}{r}{Continued on next page}\\
\endfoot
\tablebottomrule
\endlastfoot
Display-based enrichment & CD98hc & 100 & 100 \\
 & EGFR & 100 & 100 \\
 & GPC3 & 100 & 100 \\
 & PLXNA1 ectodomain & 100 & 100 \\
 & SARS-CoV-2 spike S2 & 100 & 100 \\
 & Transferrin receptor & 100 & 100 \\
 & TNFRSF23 & 100 & 100 \\
\tablemidrule
IgDesign SPR & ACVR2B & 13 & 188 \\
 & C5 & 29 & 164 \\
 & FXI & 33 & 69 \\
 & IL-36R & 28 & 137 \\
 & TNFRSF9 & 27 & 149 \\
 & TSLP & 114 & 49 \\
\end{longtable}

\subsection{Assay definitions, construct statistics, and sequence diversity}
\label{app:assay_diversity}

\paragraph{Assay and label definitions.}
The three benchmarks deliberately provide different kinds of experimental
evidence (Table~\ref{tab:assay_labels}). The display-based enrichment labels capture
selection-associated enrichment; the IgDesign labels retain only unambiguous
replicate SPR calls; and the literature/patent collection uses widely
separated direct-affinity regimes. This diversity is useful for evaluating
whether a structural filtering metric is tied to one experimental setting.

\begin{table*}[htbp]
\centering
\small
\setlength{\tabcolsep}{5pt}
\renewcommand{\arraystretch}{1.08}
\caption{\textbf{Assay and label definitions.} Display-based enrichment labels are based on
round-wise enrichment under progressively lower antigen inputs.}
\label{tab:assay_labels}
\begin{tabularx}{\textwidth}{@{}>{\raggedright\arraybackslash}p{4.0cm}>{\raggedright\arraybackslash}p{5.1cm}>{\raggedright\arraybackslash}X@{}}
\tabletoprule
Benchmark and evidence & Positive and negative definition & Exclusions \\
\tablemidrule
\textbf{Display-based enrichment}\newline Yeast or phage display with successive selection rounds and a sequencing readout. & \textbf{Positive:} up to 100 late-enriched candidates with the largest normalized-frequency log-fold change.\newline \textbf{Negative:} baseline or earliest-round sequences depleted or not enriched later, supplemented when necessary by early low-enrichment candidates. & Invalid translations, low-quality reads, exact duplicates, and over-represented near-identical families. \\
\textbf{IgDesign SPR}\newline Replicate SPR measurements released with the original inverse-folding study. & \textbf{Positive:} binding in every available replicate.\newline \textbf{Negative:} no binding in any replicate. Both classes require at least two measurements. & Positive controls, discordant replicate calls, and incomplete replicate data. \\
\textbf{Literature/patent affinity}\newline Direct $K_D$ measurements by SPR, BLI, or ITC reported in papers and patents. & \textbf{Positive:} $K_D\leq10$~nM.\newline \textbf{Negative:} $K_D\geq1{,}000$~nM. & EC$_{50}$/IC$_{50}$-only, cell-based, competition-only, ambiguous, conflicting, or construct-unresolved records. \\
\tablebottomrule
\end{tabularx}
\end{table*}

\begin{figure*}[htbp]
\centering
\includegraphics[width=\textwidth]{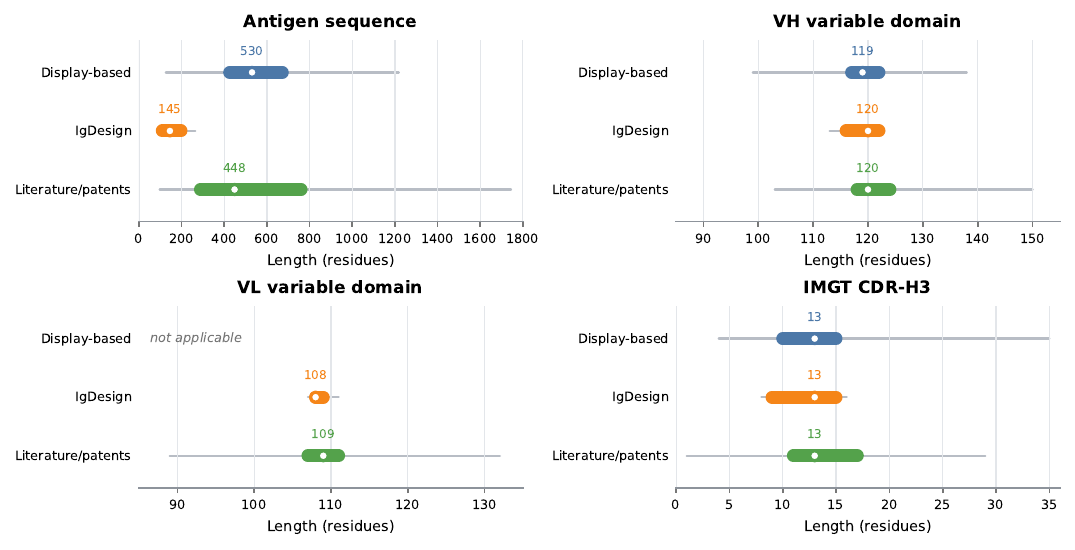}
\caption{Length profiles of the exact cofolding inputs. Thick colored segments
show the interquartile range, central dots the median, and thin lines the
observed range. Antigen, heavy-chain variable-domain (VH), light-chain
variable-domain (VL), and IMGT CDR-H3 lengths are measured in residues. VL is
absent from the heavy-chain-only display-based enrichment benchmark; IgDesign is paired
Fv, whereas the literature/patent collection contains both paired Fv and
heavy-chain-only antibodies.}
\label{fig:benchmark_lengths}
\end{figure*}

\paragraph{Construct statistics.}
Figure~\ref{fig:benchmark_lengths} summarizes the length profiles of the exact
cofolding input sequences using their median, interquartile range, and observed
range. Heavy-chain CDR3s are assigned by IMGT numbering with
ANARCI~\citep{dunbar2016anarci}. The literature/patent collection spans a
substantially broader range of antigen sequences and contains both paired Fv
and heavy-chain-only antibodies; the display-based benchmark is heavy-chain-only and IgDesign
is paired Fv by design.

\begin{figure*}[htbp]
\centering
\includegraphics[width=0.92\textwidth]{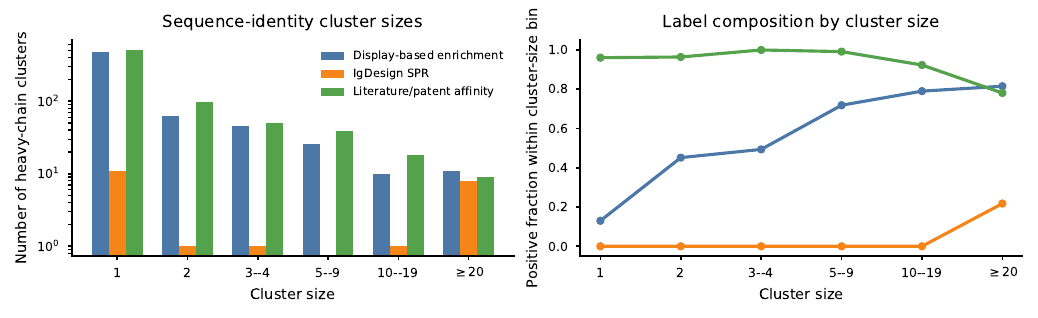}
\caption{Heavy-chain sequence-identity clusters at 90\% identity and 90\%
coverage. Left: number of clusters in each size bin. Right: candidate-level
positive fraction within clusters in each size bin.}
\label{fig:benchmark_clusters}
\end{figure*}

\paragraph{Diversity and redundancy.}
Table~\ref{tab:diversity} reports both exact sequence uniqueness and
sequence-identity clusters. Because the display-based enrichment benchmark contains only
heavy chains, heavy-chain clustering is the common antibody representation
across all datasets. We cluster heavy chains independently within each dataset
at 90\% sequence identity and 90\% coverage using MMseqs2~\citep{steinegger2017mmseqs2}; antigen constructs are clustered at 50\% identity and 80\% coverage after exact deduplication. No exact heavy-chain sequence is shared between any pair of datasets.
Figure~\ref{fig:benchmark_clusters} complements these summary statistics by
showing the full heavy-chain cluster-size spectrum and the positive fraction
within each cluster-size bin.

\begin{table*}[htbp]
\centering
\small
\setlength{\tabcolsep}{7pt}
\caption{\textbf{Diversity and redundancy of cofolding constructs.} Antigen clustering
is applied to unique construct sequences; the largest antigen cluster contains
two constructs in the literature/patent dataset and one in each other dataset.
Exact VH--VL pairs number 1,361, 1,171, and 1,608, respectively.}
\label{tab:diversity}
\begin{tabular}{@{}lrrrr@{}}
\tabletoprule
Dataset & Unique VH & \makecell{VH clusters\\(90\%/90\%)} & \makecell{Largest VH\\cluster} & \makecell{Antigen constructs /\\clusters (50\%/80\%)} \\
\tablemidrule
Display-based enrichment & 1,361 & 632 & 52 (3.7\%) & 7 / 7 \\
IgDesign SPR & 1,171 & 22 & 201 (17.2\%) & 7 / 7 \\
Literature/patent affinity & 1,453 & 723 & 38 (2.4\%) & 165 / 160 \\
\tablebottomrule
\end{tabular}
\end{table*}

\section{Experimental Details}
\label{app:experimental_details}

\subsection{Cofolding inference}
\label{app:cofolding_inference}

Every antibody--antigen pair is evaluated with the same nominal inference
budget for each co-folding predictor. MSAs are enabled, structural templates
are disabled, and the MSA and sequence inputs remain fixed across seeds and
samples. We run 10 seeds and draw 5 samples from each seed, producing at most
50 predictions per pair and predictor. Table~\ref{tab:inference_budget}
reports both the shared prediction budget and the predictor-specific parameters. 

\begin{table*}[htbp]
\centering
\small
\setlength{\tabcolsep}{7pt}
\renewcommand{\arraystretch}{1.08}
\caption{\textbf{Cofolding inference configuration for each antibody--antigen pair.} The first
five rows are shared across predictors. Remaining rows report predictor-specific
settings used at inference, including official defaults when the runner did not
override them. MSA limits denote model-input caps after MSA preprocessing; blank
cells indicate that no corresponding control applies.}
\label{tab:inference_budget}
\begin{tabularx}{\textwidth}{@{}>{\raggedright\arraybackslash}p{4.2cm}>{\centering\arraybackslash}X>{\centering\arraybackslash}X>{\centering\arraybackslash}X@{}}
\tabletoprule
Setting & AlphaFold~3 & Protenix-v2 & ESMFold2 \\
\tablemidrule
Unpaired MSA & enabled & enabled & enabled \\
Structural templates & disabled & disabled & disabled \\
Random seeds & 10 & 10 & 10 \\
Diffusion samples per seed & 5 & 5 & 5 \\
Maximum predictions per pair & 50 & 50 & 50 \\
\tablemidrule
\multicolumn{4}{@{}l}{\textit{Predictor-specific inference settings}} \\
Recycling cycles & 10 & 10 &  \\
Diffusion or sampling steps & 200 & 200 & 100 \\
Structure-update loops &  &  & 20 \\
Attention implementation & XLA &  &  \\
Numerical precision & mixed bfloat16; outputs fp32 & mixed bfloat16; confidence fp32 & mixed: ESMC bfloat16; folding fp32 \\
Language-model dropout &  &  & 0.30 \\
MSA rows retained for inference & total $\leq16{,}384$; paired $\leq600$/species & total $\leq16{,}384$; chunk size $2{,}048$ & depth $\leq1{,}024$; sequences $\leq16{,}384$ \\
MSA column-mask rate &  &  & 0.10 \\
\tablebottomrule
\end{tabularx}
\end{table*}

\paragraph{OpenDDE configuration.}
OpenDDE uses the general-purpose \texttt{opendde.pt} checkpoint, not the
antibody--antigen-specific checkpoint. We reuse the validated unpaired MSAs
used for AlphaFold~3, with no structural templates or guidance constraints.
The configuration uses seeds 101--110, five samples per seed, ten recycling
cycles, 200 diffusion steps, and mixed bfloat16 precision. We preserve
full-precision confidence outputs, including residue PAE and predicted
contact probabilities, and therefore compute RIC rather than gRIC.
The source revision (\texttt{9d0d9c9}) and full checkpoint checksum are recorded
in the released configuration. All 4,184 ensembles contain 50 predictions.

\subsection{Baseline implementation}
\label{app:baseline_details}

\paragraph{Predictor-native baselines.}
We read ranking confidence and pair-iPTM directly from the predictor output.
For complexes with multiple antibody or antigen chains, pair-iPTM is the
largest antibody--antigen entry in the reported chain-pair matrix. Antibody
pLDDT is averaged over antibody residues. Interface PAE is the mean of both
PAE directions over predicted antibody--antigen C$\beta$ contacts within
$8$~\AA{} (C$\alpha$ for glycine). A prediction without such a contact receives
the largest cross-partner PAE. PDE/gPDE is read from the corresponding
predictor output. Lower-is-better scores are sign-reversed only during
threshold evaluation.

\paragraph{Published interface scores.}
The pDockQ, pDockQ2, ipSAE, and AntiConf calculations follow their authors'
reference implementations. pDockQ uses the number of $8$~\AA{} C$\beta$ contacts
and mean interface pLDDT. pDockQ2 applies its PAE-weighted interface transform
in both partner directions and retains the larger score. Each ipSAE variant is
computed independently for every antibody--antigen chain pair, retaining the
maximum over both PAE directions and eligible chain pairs. ipSAE-d0res adapts
$d_0$ for each aligned residue, ipSAE-d0dom uses the number of residues in the
low-PAE interface footprint, and ipSAE-d0chn uses the complete chain-pair
length; all three use a PAE cutoff of $10$~\AA{}. AntiConf is
$0.3\,\mathrm{pDockQ2}_{\mathrm{Ag}}+0.7\,\mathrm{pTM}$, where the first term
uses antigen as the aligned partner.

\paragraph{Ensemble aggregation.}
Predictor-specific chain indices are converted to a common antibody and
antigen chain-role representation before scoring. Unless a method defines its
own ensemble reducer, its per-prediction values are summarized first within
each run and then across runs:
\begin{equation}
M=\operatorname*{median}_{s=1}^{10}
\left[\operatorname*{median}_{k=1}^{5} r_{s,k}\right].
\label{eq:ensemble_median}
\end{equation}
This gives every baseline the same 10-seed, 5-sample evidence budget while
preventing one unusually favorable sample from dominating a candidate score.

\subsection{Campaign-specific evaluation protocol and quality controls}
\label{app:campaign_specific_evaluation}

Table~\ref{tab:target_specific_results} evaluates the decision faced within a
single-antigen design campaign: candidates are compared only with other
candidates designed against the same antigen. We define a stratum as a
canonical (dataset, antigen) pair and require every included stratum to contain
at least 10 positives and 40 negatives. The negative-support requirement is
particularly important for the low-FPR endpoints: with fewer than 10
negatives, a single false positive changes empirical FPR by more than 10
percentage points, while 40 negatives provide four negative observations
within the R10 operating range.

The primary campaign-specific population contains the display-based enrichment and
IgDesign benchmarks. All seven display-based enrichment strata are eligible, contributing
1,400 candidates (700 positive and 700
negative). Six IgDesign strata are eligible, contributing 1,000 candidates
(244 positive and 756 negative); IL17A is excluded because it contains only
five positives. The final population therefore contains 13 strata and 2,400
candidates (944 positive and 1,456 negative). The literature/patent affinity
domain is excluded from this analysis because its eight mixed-label strata
contain only 4, 1, 1, 6, 1, 6, 2, and 1 negatives, respectively. These strata
cannot resolve the reported 10--30\% FPR operating points.

To prevent either assay domain from dominating the summary, display-based enrichment and
IgDesign each receive one half of the total weight. Within each domain, strata receive
square-root information weights. For stratum $t$, with $P_t$ positives and
$N_t$ negatives, define
\begin{equation}
I_t=\frac{2P_tN_t}{P_t+N_t},\qquad
w_t=\frac{1}{2}\,
\frac{\sqrt{I_t}}{\sum_{u:\,d(u)=d(t)}\sqrt{I_u}},
\label{eq:campaign_stratum_weights}
\end{equation}
where $d(t)\in\{\text{display-based enrichment},\text{IgDesign}\}$. Thus
$\sum_t w_t=1$. This sublinear weighting gives better-supported campaigns more
influence without allowing the largest campaign to dominate the result.

Every endpoint is computed independently within each stratum and then
aggregated as $\sum_t w_t E_t$. In particular, Recall@$U^\star$ and
Precision@$U^\star$ are weighted means of the values attained at each
stratum's own utility-maximizing threshold; they are not computed by pooling selected counts
across strata. A stratum whose utility optimum selects no candidate contributes
zero recall and precision. Average rank is computed with ties over the eight reported
endpoints after aggregation. All metrics use identical eligible candidates,
weights, score directions, threshold tie rules, and finite-score coverage
requirements.

\subsection{Threshold selection and fixed-budget evaluation}
\label{app:threshold_selection}

\paragraph{Threshold search space.}
After ensemble aggregation, metric $m$ assigns one finite score $M_i$ to each
candidate in $\mathcal D$, with larger values indicating stronger evidence.
Because a candidate is selected when $M_i\geq\sigma$, the selected set changes
only when $\sigma$ crosses an observed score. It is therefore sufficient to
search
\begin{equation}
\Sigma(m,\mathcal D)=\{+\infty\}\cup\{M_i:i\in\mathcal D\}.
\label{eq:threshold_set}
\end{equation}
This search space enumerates every distinct selection obtainable by
thresholding. The value $+\infty$ gives the empty selection; lowering
$\sigma$ through the distinct observed scores produces a nested sequence of
increasingly inclusive selections, ending with the full candidate set at the
smallest score. Thresholds strictly between adjacent score levels produce the
same selection and need not be evaluated, while candidates with equal scores
necessarily enter together. At each threshold we compute TP, FP, TN, and FN,
followed by the Wilson bounds and utility in Definitions~1--2. We choose the
threshold with maximum utility; exact utility ties favor lower FPR, then higher
recall, and finally the stricter threshold.

\paragraph{Utility-maximizing thresholds.}
Let $\mathcal D_t$ denote the candidates in campaign stratum
$t$, and let $\mathcal T$ contain the class-supported campaign strata defined
in Appendix~\ref{app:campaign_specific_evaluation}. The portfolio-wide and
campaign-specific utilities are
\begin{equation}
\begin{aligned}
U_{\mathrm{portfolio}}^\star(m)
&=\max_{\sigma\in\Sigma(m,\mathcal D)}U(m,\sigma;\mathcal D),\\
U_{\mathrm{campaign}}^\star(m)
&=\sum_{t\in\mathcal T}w_t
\max_{\sigma_t\in\Sigma(m,\mathcal D_t)}
U(m,\sigma_t;\mathcal D_t).
\end{aligned}
\label{eq:global_target_utility}
\end{equation}
Portfolio-wide filtering pools all candidates and uses one
shared threshold, testing cross-antigen score calibration. Campaign-specific
filtering repeats the threshold search within each stratum, testing how well a
metric ranks antibodies when the antigen is fixed. Portfolio-wide precision
is evaluated at the maximizing shared threshold $\widehat\sigma$. If
$\widehat\sigma_t$ maximizes utility in campaign $t$, campaign-specific
precision is
\begin{equation}
P_{\mathrm{campaign}}@U^\star
=\sum_{t\in\mathcal T}w_t P_t(\widehat\sigma_t),
\label{eq:precision_at_target_utility}
\end{equation}
where $P_t(\widehat\sigma_t)$ is set to zero when the
utility-maximizing threshold selects no candidate.

\paragraph{Sensitivity to the FPR penalty.}
We repeated the threshold searches with $f_{\mathrm{safe}}=0.10$ fixed and
$f_{\mathrm{severe}}$ increased from $0.30$ to $0.40$ and $0.50$. RIC
(gRIC for ESMFold2) retained the highest utility in portfolio-wide filtering,
and NAC retained the highest utility in campaign-specific filtering, for all
three predictors at both alternative settings. Relaxing the penalty generally
increased recall but also admitted more false positives. For example, NAC's
weighted FPR at its utility-maximizing thresholds rose from approximately
8\% to 15\% when $f_{\mathrm{severe}}$ increased from $0.30$ to $0.50$.
The main analysis retains $f_{\mathrm{severe}}=0.30$ to emphasize low-FPR
filtering.

\paragraph{Recall at a fixed false-positive rate.}
This evaluation asks how many binders can be retained while allowing at most a
specified fraction $\alpha$ of non-binders through the filter. Among all
attainable thresholds satisfying that empirical FPR limit, we report the
largest recall. For portfolio-wide filtering this is
\begin{equation}
\operatorname{Recall}_{\mathrm{portfolio}}@\operatorname{FPR}\alpha
=\max_{\substack{\sigma\in\Sigma(m,\mathcal D)\\F(\sigma;\mathcal D)\leq\alpha}}
R(\sigma;\mathcal D).
\label{eq:global_recall_at_fpr}
\end{equation}
Campaign-specific filtering performs the same constrained
search separately within each stratum and then applies the fixed weights in
Equation~\ref{eq:campaign_stratum_weights}:
\begin{equation}
\operatorname{Recall}_{\mathrm{campaign}}@\operatorname{FPR}\alpha
=\sum_{t\in\mathcal T}w_t
\max_{\substack{\sigma_t\in\Sigma(m,\mathcal D_t)\\
F_t(\sigma_t;\mathcal D_t)\leq\alpha}}
R_t(\sigma_t;\mathcal D_t).
\label{eq:target_recall_at_fpr}
\end{equation}
We use $\alpha\in\{0.10,0.20,0.30\}$. Unlike utility, which uses Wilson bounds
to account for finite-sample uncertainty, this operating-point analysis
constrains the observed FPR directly. The empty selection has $F=R=0$ and is
always feasible. If every nonempty selection exceeds the FPR limit, the
reported recall is therefore zero; in campaign-specific
evaluation, that stratum contributes zero with its prespecified weight.

\paragraph{Recall at a fixed testing budget.}
Fixed-FPR evaluation constrains screening errors, whereas fixed-budget
evaluation constrains the number of candidates that can be tested. For a
testing fraction $q$, let $K_q=\lfloor qN\rfloor$ and retain the $K_q$
highest-ranked candidates. This is a top-$K_q$ decision rather than a search
for the utility-maximizing threshold. If the boundary cuts through a tied
score group, let $\mathcal H_q$ contain candidates strictly above the boundary
and $\mathcal E_q$ those tied at the boundary, and define
\begin{equation}
\rho_q=\frac{K_q-|\mathcal H_q|}{|\mathcal E_q|}.
\label{eq:budget_tie_fraction}
\end{equation}
The expected number of retained positives under label-independent random
tie-breaking is
\begin{equation}
\operatorname{TP}_q
=\sum_{i\in\mathcal H_q}Y_i
+\rho_q\sum_{i\in\mathcal E_q}Y_i,
\qquad
\operatorname{Recall}@Q=q=\frac{\operatorname{TP}_q}{\sum_iY_i}.
\label{eq:recall_at_budget}
\end{equation}
We report $q\in\{0.05,0.20\}$. In portfolio-wide filtering,
all candidates compete for a single testing budget. In campaign-specific
filtering, the same fraction is retained independently within each stratum and
recall is aggregated with the fixed campaign weights. When tied scores cross
the budget boundary, the expression above is
the expected recall under label-independent random tie-breaking, preventing
file order from determining the result. Recall is defined as zero when
$K_q=0$.

\section{RIC Implementation Details}
\label{app:ric_implementation}

\paragraph{Contact forecast and PAE agreement.}
For each reference run $r$, we average both directions of predicted contact
probability across its samples:
\begin{equation}
\bar C_r(a,b)
=\frac{1}{|\mathcal K_r|}\sum_{j\in\mathcal K_r}
\frac{C^{\mathrm{Ab}\to\mathrm{Ag}}_{rj}(a,b)
+C^{\mathrm{Ag}\to\mathrm{Ab}}_{rj}(b,a)}{2},
\qquad h_r(a)=\max_{b\in\mathcal G}\bar C_r(a,b).
\label{eq:ric_contact_support}
\end{equation}
For sample $j$ in run $r$ and sample $k$ in another run $t$, their
bidirectional PAE agreement is
\begin{equation}
\begin{aligned}
A_{rtjk}(a,b)=\frac{1}{2}\Bigg[&
\frac{\tau}{\tau+P^{\mathrm{Ab}\to\mathrm{Ag}}_{rj}(a,b)
+P^{\mathrm{Ab}\to\mathrm{Ag}}_{tk}(a,b)}\\
&+\frac{\tau}{\tau+P^{\mathrm{Ag}\to\mathrm{Ab}}_{rj}(b,a)
+P^{\mathrm{Ag}\to\mathrm{Ab}}_{tk}(b,a)}\Bigg],
\qquad \tau=16\ \text{\AA}.
\end{aligned}
\label{eq:ric_pae_agreement}
\end{equation}

An exploratory sensitivity analysis compared PAE scales of $4$, $8$, and $16$~\AA. The $16$~\AA\ scale performed best overall, although $8$~\AA\ led on some subset-level measures; the differences were generally modest. We therefore used $16$~\AA\ for all predictors.

\paragraph{Tie-aware rank weights.}
For a reference run $r$ and CDR residue $a$, antigen residues are ranked by
$\bar C_r(a,b)$.  With rank budget
$m=\min(8,\lfloor |\mathcal G|/2\rfloor)$, the high-contact weights
$w^H_{ra}$ allocate unit mass to ranks $1,\ldots,m$, and the comparison
weights $w^L_{ra}$ allocate unit mass to ranks $m+1,\ldots,2m$.  If equal
contact values cross either boundary, the total mass assigned to the affected
ranks is divided equally among all residues in the tied block.  The resulting
weights therefore do not depend on the arbitrary order of tied residues.

\paragraph{Short antigens.}
When $2\leq |\mathcal G|<16$, the adaptive definition of $m$ contracts both
rank regions so that they fit within the available antigen residues; if the
antigen length is odd, the final unpaired residue is unused.  For the limiting
case $|\mathcal G|=1$, the sole residue receives unit weight in both regions,
and the local contrast $\Delta_{rtjk}(a)$ in Equation~\ref{eq:ric_score} is exactly zero.

\paragraph{Zero contact support.}
If $h_r(a)=0$ for every CDR residue in a reference run, the normalized weights
$h_r(a)/\sum_{a'}h_r(a')$ are not evaluated and all contributions from that
reference run are defined as zero. This remains a valid zero-valued observation
in the median hierarchy rather than being treated as missing data.

\paragraph{Geometric-contact approximation.}
For ESMFold2, which does not provide contact probabilities, gRIC replaces both
contact directions in Equation~\ref{eq:ric_contact_support} with the same
binary contact map: antibody and antigen residue representative atoms are in
contact when their predicted distance is less than $8$~\AA. The PAE agreement,
rank weights, contact-support weighting, and median aggregation remain those
of RIC.

\paragraph{Chain mapping and ensemble completeness.}
The antibody and antigen scopes are defined by the chain roles supplied for
each candidate.  A single-chain antibody is scored using that chain alone; no
placeholder light-chain residues are introduced.  Every declared chain must
map one-to-one to the predictor tokens and input sequence.  Across the
ensemble, all predictions must preserve the same ordered antibody residues,
antigen residues, and CDR annotation, and all contact and PAE tensors must have
matching finite dimensions.  The experiments in this paper require the full
10-run, 5-sample ensemble.  A missing prediction, a missing declared chain, or
an inconsistent residue mapping invalidates the candidate score. We do not
aggregate predictions from an incomplete ensemble.

\section{Related Work}
\label{sec:related}

\paragraph{Computational antibody and binder design.}
Modern antibody and protein-binder design spans backbone generation, joint
sequence--structure generation, inverse folding, and predictor-guided sequence
optimization. RFantibody adapts diffusion-based structure generation to
antibody scaffolds and CDR loops, whereas BindCraft optimizes binder sequences
through a structure-prediction network
\citep{bennett2026rfantibody,pacesa2025bindcraft}. Chai-2, BoltzGen, PPIFlow,
and ODesign broaden this space toward zero-shot antibody design, multimodal
binder generation, and joint modeling of sequence, structure, and molecular
interactions \citep{chai22025,stark2025boltzgen,yu2026ppiflow,zhang2025odesign}.
Despite their architectural differences, these methods share a practical
dependence on predicted complex structures and post-generation filtering to
select candidates for experiment.

\paragraph{Co-folding models for antibody--antigen complexes.}
Modern co-folding methods have progressively expanded from protein-complex
prediction to general all-atom modeling of biomolecular interactions.
AlphaFold-Multimer established joint prediction of multichain protein
assemblies and introduced interface-aware confidence through ipTM
\citep{evans2022multimer}. More recent all-atom systems, including
AlphaFold~3, Protenix-v2, Chai-1, Boltz-2, and OpenDDE, extend this paradigm to
heterogeneous biomolecular complexes and provide increasingly rich structural,
pairwise, and interface-level outputs
\citep{abramson2024alphafold3,zhang2026protenix,chai12024,
passaro2025boltz2,opendde2026}. Within this family, Boltz-2 further couples
structure prediction with binding-affinity prediction
\citep{passaro2025boltz2}. In parallel, ESMFold2 follows a
protein-language-model-centered route, using learned evolutionary
representations for efficient prediction of
protein structures and complexes \citep{candido2026esmfold2}. Together, these
models provide increasingly diverse predictions of antibody--antigen geometry
and confidence, creating a rich source of structural evidence for downstream
candidate prioritization.

\paragraph{Structural confidence metrics.}
The native ipTM score summarizes confidence in the relative placement of
chains, but its whole-chain normalization can dilute a localized interface.
pDockQ combines interface pLDDT with the number of predicted contacts to
estimate DockQ, and pDockQ2 incorporates interchain PAE to score individual
interfaces in multimeric complexes
\citep{bryant2022pdockq,zhu2023pdockq2}. ipSAE restricts scoring to low-PAE
interchain residue pairs and varies the TM normalization scale with the
supported interface or chain size \citep{dunbrack2025ipsae}. AntiConf further combines
confidence measures for antibody--antigen complex assessment
\citep{unsal2026anticonf}. These advances improve estimates of predicted
interface quality; our work asks the complementary decision question of how
structural confidence should be aggregated to retain true binders while
controlling false positives in in silico antibody filtering.

\end{document}